\documentclass{article}

\title{A Survey of Intrusion Detection Systems Leveraging Host Data} 
\author{
  Glass-Vanderlan, Tarrah R.\\
  \texttt{glassvandetr@ornl.gov}
  \and
  Iannacone, Michael D.\\
  \texttt{iannaconemd@ornl.gov}
    \and
  Vincent, Maria S.\\
  \texttt{vincentms@ornl.gov}
    \and
  Chen, Qian (Guenevere)\\
  \texttt{geunevereqian@utsa.edu}
    \and
  Bridges, Robert A.\\
  \texttt{bridgesra@ornl.gov}
}

\usepackage[margin=1.2in]{geometry}

\usepackage{booktabs} 
\usepackage{hyperref} 
\usepackage{lipsum}
\usepackage{courier} 
\usepackage{tabularx}
\usepackage{graphicx}
\graphicspath{ {./images/} } 
\usepackage{pdflscape}
\usepackage{makecell}
\usepackage[numbers]{natbib} 
\usepackage{enumitem} 
\usepackage[flushleft]{threeparttable}
\usepackage{multirow}

\usepackage{todonotes} 
\usepackage{wrapfig}
\usepackage[ruled]{algorithm2e} 
\usepackage{xcolor,colortbl}
\usepackage{ragged2e} 

\usepackage{float}

\definecolor{LightCyan}{RGB}{112,172,203}
\definecolor{LightGrey}{RGB}{238,238,238}

\SetAlFnt{\small}
\SetAlCapFnt{\small}
\SetAlCapNameFnt{\small}
\SetAlCapHSkip{0pt}
\IncMargin{-\parindent}
\usepackage{soul}
\usepackage{color}

\begin{document}
\maketitle

\begin{abstract}

This survey focuses on intrusion detection systems (IDS) that  leverage host-based data sources for detecting attacks on enterprise network. 
The host-based IDS (HIDS) literature is organized by the input data source, presenting targeted sub-surveys of HIDS research leveraging system logs, audit data, Windows Registry, file systems, and program analysis. 
While system calls are generally included in audit data, several publicly available system call datasets have spawned a flurry of IDS research on this topic, which merits a separate section. 
Similarly, a section surveying algorithmic developments that are applicable to HIDS but tested on network data sets is included, as this is a large and growing area of applicable literature. 
To accommodate current researchers, a supplementary section giving descriptions of publicly available datasets is included, outlining their characteristics and shortcomings when used for IDS evaluation. 
Related surveys are organized and described. 
All sections are accompanied by tables concisely organizing the literature and datasets discussed. 
Finally, challenges, trends, and broader observations are throughout the survey and in the conclusion along with future directions of IDS research.

\end{abstract}


\thanks{This manuscript has been authored by UT-Battelle, LLC, under contract DE-AC05-00OR22725 with the US Department of Energy (DOE). The US government retains and the publisher, by accepting the article for publication, acknowledges that the US government retains a nonexclusive, paid-up, irrevocable, worldwide license to publish or reproduce the published form of this manuscript, or allow others to do so, for US government purposes. DOE will provide public access to these results of federally sponsored research in accordance with the DOE Public Access Plan (http://energy.gov/downloads/doe-public-access-plan).}
\maketitle

\section{Introduction} 
Intrusion detection research began in 1972, when James Anderson published a United States Air Force report discussing the need to detect security breaches of computing systems~\cite{anderson1972computer}. 
Manual investigations of logs and audit data were widely adopted by computer security operators (or system administrators) in the early age of IT technology, yet IDSs that fully depended on experienced security experts could not meet the new requirements of the developing computing technology.  
In response, automated IDS research emerged\textemdash Anderson's 1980 work~\cite{anderson1980computer} focused on automating IDS by isolating abnormal behavior in system's audit data, and work of Denning and Neumann~\cite{denning1985requirements}  developed the first real-time detection system based on expert-written rules in 1985. 
This early research laid the groundwork for modern \textit{intrusion detection}, comprised of manual techniques, algorithms, and commercial products all geared towards one thing, continual monitoring of computing assets for signs of compromise.~\cite{bruneau2001history, kemmerer2002intrusion}

Increasingly over the last 30 years, networked computing systems have emerged as ubiquitous assets on which state, personal, and industrial infrastructure critically depend. 
Moreover, the threat of cyber security breaches has risen, with adversaries now fueled by a profitable underground cyber-crime economy and nation-state ambitions~\cite{alazab2012cybercrime, sood2013cybercrime}. 
Consequently, breaches ranging from personal computers to large enterprises and governmental networks are now commonly reported, and governmental assistance, in terms of IDS tutorials, guidelines, and case studies have resulted \cite{scarfone2007guide}. 
Through the authors' ongoing collaborations with multiple security operations, we note that many operations now have widespread collection and query capabilities for logs and alerts.  
Yet, detection in practice has focused on signature and rule-based detection, often at the network or file levels, complemented by manual analysis of logs. \cite{chen2017automated, bridges2017setting}  
These rule-based IDSs are accurate for detecting known system cyber attacks but cannot identify unknown, novel, or polymorphic cyber threats. 
In addition, their computational overheads (i.e., time, CPU, and memory costs) are usually high. 
This has motivated parallel developments in the research literature for a wide variety of automated, fast, and efficient IDSs. 
From expert-crafted rules to sophisticated statistical learning algorithms, publications explore and push detection accuracy metrics and performance on a variety of data sources and locations within the network (see Sections \ref{sec:sys-log-audit}-\ref{sec:other-data} below). 

\subsection{IDS Components, Types, and Challenges}
In general, all intrusion detection systems (IDSs) have three main components.
\begin{itemize}[leftmargin=.7cm]
\item \textit{Data collection:}   They ingest one or many data types e.g., system calls or network flows. 
\item  \textit{Conversion to select features:} Some predefined unit of data, e.g., system calls in a process or flows in a time window, is represented as a list of attributes, called a feature vector. 
\item \textit{Decision engine:} An algorithm or heuristic to decide if the given data, as represented as a feature vector, is believed to be an attack or not.
\end{itemize}
Common research for IDS development involves testing supervised classifiers and unsupervised anomaly or one-class detectors as the decision engine algorithm. 
The decision engine can then be configured to inform a user or some automated response system.

The decision engine can be categorized as \textit{misuse}, \textit{anomaly} or a \textit{hybrid} detector.  
Misuse intrusion detection uses predefined attack patterns, e.g., signatures of known malware or expert-crafted rules, to flag matching events. 
Consequently, zero-day attacks, i.e., novel attacks or attacks exploiting previously unknown vulnerabilities, generally bypass misused detection algorithms. 
Misuse detection systems dominate IDS use in practice, as they have been the main focus of commercially-driven detection products. 
Host-based anti-virus such as McAfee\footnote{ See \url{www.mcafee.com}.} and Kasperski\footnote{ See \url{www.kaspersky.com}.}, and network-level rule-based systems such as Snort\footnote{ See \url{www.snort.com}.} are examples of very popular misuse detection systems. 
Generally, the first line of defense, misuse detection relies on an database of attack signatures, which is generally large, constantly growing, can be cumbersome to use efficiently and necessitates regular updates. 
On the other hand, misuse detection generally realizes a relatively low number of false positives, but a high number of false negatives.

In anomaly detection, a description of normal or expected behavior is learned from observations and a sufficient deviation from this normal profile is flagged as a potential attack; thus, the detection of never-before-seen attack patterns is possible. 
Anomaly detection systems that update in near real time can evolve models with the slowly changing system~\cite{ferragut2012new, harshaw2016graphprints}. 
The primary downside of anomaly detection is detection accuracy, most notably, that these techniques suffer from higher quantities of false alarms. 
Moreover, attacks can hide in the noise floor of ambient data if training data (from which normal behavior is learned) exhibits large variance. 
Similarly, if attacks are present in training data, detectors will potentially be trained to regard such behavior as normal~\cite{sommer2010outside}.

Hybrid systems are also often studied; these take into account previous knowledge but seek to generalize to unseen data. 
For example, systems are proposed that seek to complement misuse detection with anomaly detection, using them in tandem \cite{veeramachaneni2016ai}. 
When datasets with labeled intrusions are available, research will often experiment with combinations of feature selection and supervised learning algorithms.   
Supervised learning classifiers are generally less rigid than traditional misuse detection systems, as they  are trained to generalize previously seen attack and non-attack examples. 
Supervised learners can often classify anomalies as attacks.

Feature selection is influential  on both accuracy and performance of 
IDSs classifiers. 
In many applications, the number of features can grow to enormous quantities, but as feature vectors gain length so does the computational complexity, quantity of training data, and time needed for both training and inference. 
Additionally, poor features both decrease performance and add noise, reducing accuracy of the classifier while contributing to expense. 
To combat these factors, methods of dimension reduction seek to identify redundancy and find correlations in features, thereby reducing the number of features without losing information.  
This careful choice of features is the focus of many detection efforts. 
Where progressions of research built on the same datasets exist (e.g., see Sections \ref{sec:system-calls} and \ref{sec:other-data}), research generally trends from using raw data as features, to considering cost-to-accuracy benefits of various hand-crafted features, to data-driven techniques for dimension reduction and feature selection/creation. 

Biased classes, in our case where non-attack data is in far more abundance than attack data, are a perennial problem for classifiers and a looming issue in the intersection of machine learning and intrusion detection. 
Much research seeks to use hybrid methods, ensembles, and advanced feature selection algorithms to circumvent the problem. 

While incomplete training data (in particular, not having representative attack data available) is an issue for misuse supervised detectors, noisy training data is a common problem especially for anomaly detectors, which often characterize normal data from a history of the data. 
Most notably, if unknown attacks exist in training data, the detector may regard similar future attacks as normal. 
Robust statistical methods have been used to discard outliers when fitting anomaly detection models, which can help address these problems. 
As our survey is organized by data source, unique approaches to these challenges are pointed out in the sections in which they occur.

\subsection{IDS Location}
\label{sec:location}

An IDS is most often categorized based on the information source utilized by the IDS and its position within the network architecture. 
Since an IDS's capabilities depend largely on what data it has available~\cite{modi2013survey}, location is a critical architectural decision. 
This can be viewed most coarsly as network-based IDS (NIDS) versus host-based IDS (HIDS).  
Host-based IDSs are generally a software component located on the system being monitored, and typically monitor a single system.  
This gives HIDS excellent visibility into the system state, but poor isolation from the system, meaning that an attacker with access to the system can either mislead or disable the HIDS.  
Additionally, host-based data is often context-rich, allowing deeper understanding of processes and activities, but comes with added costs of requiring access to the host, configuration of distributed clients, and often requires collecting and managing potentially large and sensitive datasets from these hosts. 
Network IDSs are generally physically separate devices, located on the network ``upstream'' of the system being monitored, and they generally monitor many separate systems on a common network.  
The NIDS is often completely transparent to the systems being monitored, which provides good isolation and makes NIDSs much less susceptible to any interference from an attacker.  
However, these systems have little or no information available about the internal state of the systems they are monitoring, which can make detection more difficult.

Hybrid and distributed IDSs (DIDS) combine information from multiple sources into one system. Hybrid IDSs combine both host-based and network-based data, generally with the goal of achieving more complete visibility of a host.  
Distributed IDSs combine data from multiple sensor locations into a combined decision-making or combined alerting process; this may use information from host-based sensors, network-based sensors, or both.

The use of virtualized resources, e.g., cloud computing environments, provide opportunities for monitoring virtualized hosts from different locations, with trade offs in visibility and capabilities. 
For example, traditional IDSs can reside inside the virtualized host, or one can gain isolation from infections or compromises of the host, at the cost of poorer visibility into the host, by monitoring host data at the hypervisor-level to perform detection of one or many guest Operating Systems (OSs). 
Virtual machine monitor (VMM) IDSs involve monitoring a virtual machine's (VM) OS (or in some cases, its applications or services) from a logically external location on the same physical machine.  
Several of the above IDS techniques have been utilized within cloud environments. 
These cloud IDSs can include network-based data, host-based data, or both, and, while cloud infrastructure relies heavily on VMs, these systems do not necessarily include the same techniques as VMM IDSs. 

Among HIDS, most systems can be categorized as either a program-level or OS-level IDS. 
Program-level IDSs focus on monitoring a single application, using information such as source code, byte code, system calls invoked, static or dynamic control flow, and other information on the application's state. 
Much of this relies significantly on research in related topics, such as vulnerability detection and malware analysis/detection, but here we focus on works that take these techniques and apply them to detecting intrusions or anomalies in applications at run-time.

OS-level IDSs monitor the overall system state, and may monitor the combined behavior of all processes, to distinguish between normal and abnormal behavior at the OS level. 
This can involve collecting data from system logs, Windows Registry data, system calls invoked, file system monitoring, or other sources.
System calls have been well utilized for the detection of normal and anomalous behavior. 
While sequences of system calls for a single application can be used in a program-level IDS, these are often combined, by monitoring all system calls of all processes, for use in an OS-level IDS.
System call traces, the sequence of system calls of a given process, are used to find repeated patterns of system calls, enabling anomaly detection and misuse detection during execution.~\cite{kosoresow1997intrusion}.

Finally, we note that side-channel detection\textemdash detectors leveraging physical characteristics such as power consumption, electromagnetic radiation, vibrations, timings\textemdash has gained traction in cyber-physical IDS research~\cite{gonzalez2011, choi2018identifying, clark2013wattsupdoc, jimenez2017malware} but is a growing area of research for traditional computers using host-level (albeit physical) data \cite{dawson2018rootkit, jimenez2017malware}. 
For most of these works, the primary advantage is the detectors' physical separation from the host, which prevents software intrusions from tampering with the detector, and the fact that malicious changes to host necessarily induces physical changes from normal behavior. 
These researches are considered out of scope for the current survey.

\subsection{Scope \& Organization}
This survey focuses on HIDSs, and attempts to capture and organize the variety of data sources used, methods tested, and general trends in the HIDS research. 
Because of the large volume of work in this area, we cannot comprehensively cover all relevant works; we prioritize a broad coverage each host-based data source and its use for intrusion detection, and we discuss research trends over time for each of these data sources. 
Algorithmic research that is developed for NIDS, but portable to host data sources, is isolated and included. 
While VMM-IDSs and DIDSs leverage host-level data, their contributions generally focus on new architectures, instead of the analysis itself, and for this reason they are not included.  

Section \ref{sec:related} describes several other IDS surveys. 
The following sections are sub-surveys of HIDS research,  each  for a different input data type. 
Section \ref{sec:sys-log-audit} focuses on system logs and audit data. 
While system calls are considered audit data, a flurry of targeted detection research brought on by labeled data sets merited their own Section, \ref{sec:system-calls}. 
Section~\ref{sec:registry} gives, to the best of our knowledge, a comprehensive description of the few IDS works leveraging Windows Registry data. 
Section~\ref{sec:filesystem} reviews works leveraging file system monitoring for identifying malicious files, while   Section~\ref{sec:program} discusses a few works that leverage information about processes or stored binaries on a host for detection. 
A large body of research focuses more on algorithmic development for detection than on specific data source applications; in particular, many of these works test the proposed methods on KDD- and DARPA-related network datasets, but the algorithms are applicable to host-based data sets. 
Many of these works are captured in Section~\ref{sec:other-data}. 
Our final section gives conclusions. 
To assist current research, the publicly available datasets and databases referenced in the literature for IDS validation are collected in \hyperref[sec:datasets]{Supplemental Materials}.


All sections are accompanied by one or many tables, itemizing the discussed references and presenting their key characteristics for comparison. 
By organizing the literature by data source, we hope that current researchers (1) quickly see the panorama of data sources available for HIDS research, and (2) for a given data source of interest, elicit progressions in the literature and identify gaps, trends, or novel directions for future contributions.

\section{Related Surveys}
\label{sec:related}
Several surveys provide discussions on existing IDS research, and we review the recent, related ones in this section.  
Table~\ref{tab:survey} presents a quick comparison, as it details many discussion topics and attributes of the surveys.


\begin {table}[h!]
\centering
\small
\caption {Existing IDS Survey Comparison} 
\label{tab:survey} 
\begin{tabular}{p{3.7cm} p{3.3cm} p{1.5cm} p{1.2cm} p{1.2cm} p{1.4cm} }
    \toprule
    Survey & IDS Types & IDS \phantom{aaa} Location & Datasets & Data Source & Feature \phantom{a} Selection   \\
    \midrule
    This Survey & HIDS, NIDS 
    &  Discussed & Discussed & Discussed & Discussed \\
    Axelsson `00 \cite{axelsson2000intrusion} & HIDS, NIDS& NA & NA & Discussed & Discussed \\
    Patcha \& Park `07 \cite{patcha2007overview} & HIDS, NIDS  & NA & Discussed & Discussed & Discussed \\
    Kabiri \& Ghorbani `05 \cite{kabiri2005research} & NIDS & NA & NA & NA & Discussed \\
    Lazarevic et al. `05 \cite{lazarevic2005intrusion} & HIDS, NIDS, DIDS  & NA & NA & Discussed & NA \\
    Sabahi et al. `08 \cite{sabahi2008intrusion} & HIDS, NIDS & Discussed & Discussed & Discussed & NA \\
    Mehmood et al. `13 \cite{mehmood2013intrusion} & HIDS, NIDS, DIDS, VMM  & Discussed & NA & Discussed & Discussed \\
    Liao et al. `13 \cite{liao2013intrusion} & HIDS, NIDS,  DIDS, VMM  & NA & NA & Discussed & NA \\
    Modi et al. `13 \cite{modi2013survey} & HIDS, NIDS, DIDS, VMM  & Discussed & NA & NA & NA \\
    Kumar \& Gohil `15 \cite{kumar2015survey} & HIDS, NIDS, DIDS  & Discussed & NA & NA & NA \\
    Kahn et al. `16 \cite{khan2016survey} & HIDS, NIDS  & Discussed & NA & NA & NA \\
    Chiba et al. `16 \cite{chiba2016survey} & HIDS, NIDS, DIDS, VMM  & Discussed & NA & NA & NA \\
    Buczak \& Guven `16 \cite{buczak2016survey} & NIDS  & NA & Discussed & Discussed & NA \\
    Mishra et al. `17 \cite{mishra2017intrusion} & VMM  & Discussed & NA & NA & NA \\
    \bottomrule
\end{tabular}
\end{table}

Axelsson~\cite{axelsson2000intrusion} surveys anomaly and misuse papers pre-2000 and organizes the research by sorting on the proposed problem's level of difficultly.   

Patcha \& Park~\cite{patcha2007overview} perform an in-depth review of anomaly and hybrid based intrusion detection papers spanning from 2000-2006. The papers are organized based on the classification algorithm used and discussed in terms of existing challenges, such as high false alarm rate. Additionally, numerous open challenges, such as failure to scale to gigabit speeds, are discussed.

Kabiri \& Ghorbani~\cite{kabiri2005research} primarily focus on NIDS, but discuss the importance of feature selection with respect to dimension reduction, importance of the features, and their relation to one another in the feature space, which is neglected as its own topic in other surveys. 

Lazarevic et al.~\cite{lazarevic2005intrusion} extensively cover attack types and categorizes them into classes. A generic architecture is defined for an IDS. The survey provides an overview on IDS taxomony and discusses information sources, including system commands, accounting, and logs as well as security audit processing. However, user-level logs, process profiling, file system, registry, raw pages/introspection are excluded. 

Sabahi et al.~\cite{sabahi2008intrusion} provide a very brief survey of different IDS systems including HIDS, NIDS, and DIDS, covering data sources used to conduct detection and detection methods, such as misuse detection, protocol analysis and anomaly detection. They mention that detection can be conducted online or offline and provide examples of both centralized and distributed architectures.

Mehmood et al.~\cite{mehmood2013intrusion} provide an overview of different intrusions for cloud-based systems and analyze several existing cloud-based IDSs with respect to their type, positioning, detection time, detection technique, data source, and attacks detection capabilities. The analysis also provides limitations of each technique to evaluate whether each fulfills the security requirements of the cloud computing environment.

Liao et al.~\cite{liao2013intrusion} present a comprehensive survey of IDSs concentrating on signature-based, behavior-based, and specification-based methods. These detection methods are further divided into ``statistics-based, pattern-based, rule-based, state-based, and heuristic-based" approaches.  

Modi et al.~\cite{modi2013survey} offer recommendations for IDS/IPS placement within cloud environments to reach common security goals in next-generation networks. 


Kumar \& Gohil~\cite{kumar2015survey} discuss traditional attack types and analysis techniques used for HIDSs, NIDSs, and DIDSs. 

Khan et al.~\cite{khan2016survey} briefly discuss HIDSs and NIDSs, and discuss their architecture and applicability, as well as highlighting shortcomings, such as the high communication and computational overhead of some approaches. A parametric comparison of the threats being faced by cloud platforms is performed, which incorporates a discussion of how various intrusion detection and prevention frameworks can apply to various common security issues.

Chiba et al.~\cite{chiba2016survey} discuss cloud-based IDSs and analyze the systems based on their various types, positions, detection type, and data source. Strengths and weaknesses are discussed to determine the IDSs validity in a cloud computing environment.

Buczak and Guven~\cite{buczak2016survey} provide a survey of machine learning approaches for IDSs. 
Their work provides brief descriptions of important algorithms, including a table with algorithmic time complexity. 
This non-comprehensive survey primarily includes examples of NIDS research, with selection criteria for influential works to include examples of each classification algorithm used for an IDS. 
Similarly, a few data sources, e.g. packets, are discussed in detail, along with several open source datasets. 
 
Mishra et al.~\cite{mishra2017intrusion} heavily focus on virtual machine introspection (VMI) and hypervisor introspection (HVI) as IDS techniques, and compares cloud security with network security. 
Cloud-specific threats and vulnerabilities are discussed via an attack taxonomy. Challenges are briefly discussed including availability of data sets, IDS position, performance, and IDS limitations. Parallel programming and the usage of GPUs are mentioned for performance improvement.

This survey provides an in-depth discussion of IDS work that leverages host-based data sources for attack detection. 
We organize works based on their data source with the goal of giving the interested reader a panoramic view of the different avenues for detection. Furthermore, this work provides an in-depth introduction to available host-level data sources, and discusses and their uses and limitations. 
Works that focus on algorithmic development, but are tested on network-level data, are included where they apply to HIDS.
Additionally, a supplementary includes a list of publicly available datasets used in the research literature for evaluating IDSs, outlining their characteristics and shortfalls. 

\section{System Log and System Audit Data IDSs}
\label{sec:sys-log-audit} 

Log files are a collection of system-generated records that detail the sequence of events of a server, an OS, or an application. 
The log files are processed or stored for various analyses or forensics. 
Most programs and applications generate separate individual log files, associated with activities conducted by those programs' processes. 
As an example, a system log file is usually associated with records produced by the OS, including but not limited to warnings, errors, and system failures. 
Individual applications may produce log files associated with user sessions containing login time, authentication result, user-program interactions, etc. 
While an OS-produced log file is considered a system log file, files produced by individual applications or users are considered audit data. 
Examples include successful and failed authentication logs, system calls, or user command logs.  

Since such data sources document the sequence of events of the system or programs, they are a promising resource for detecting intrusions, as an HIDS can leverage the data to profile behavior of an individual user or system.
Conversely, the downside to high fidelity audit data is the collection cost. 
Below we survey literature leveraging system logs and audit data for intrusion detection. 
We note that system calls can be considered a subset of audit data, but because there is a rich progression of research which considers them independently, system-call-based IDSs merit their own section, Section \ref{sec:system-calls}. 

\subsection{System Logs IDSs} 
\label{sec:sys-log} 
With any IDS, the goal is to perform in a cost-effective, adaptable, intelligent, and real-time manner. 

This is especially challenging when analyzing system logs, which can be CPU intensive and typically requires human expertise. 
System log analysis requires that all performed actions by the OS be stored, and then feature extraction and classification can then be performed. 

The following papers focus on analyzing system logs and various ways to achieve this goal. 
We break them into two subcategories\textemdash those focusing on detection accuracy, and those focusing on IDS architecture. 

\begin{wraptable}[9]{r}{.81\textwidth}
\small
\vspace{-.7cm}
\caption {HIDS with System Logs} \label{tab:systemlogs} 
\begin{tabular}{ccccc}
    \toprule
IDS Reference & Technique & Dataset & Classifier & Learning \\
   \midrule
Reuning `04 \cite{reuning2004applying} & Anomaly & DARPA99 & TF-IDF & Supervised \\
Guan et al. `05 \cite{guan2005collaborative} & Anomaly & NA  & NN & Supervised \\
Zhaojun \& Chao `10 \cite{zhaojun2010statistic} & Anomaly & NA & NN & Supervised \\
Tchakoucht et al. `15 \cite{tchakoucht2015behavioral} & Anomaly & Simulation & Clustering & Unsupervised \\
Wang \& Zhu `17 \cite{wang2017centralized} & Anomaly & KDD99 & C5.0 DT & Supervised \\
    \bottomrule
\end{tabular}
\end{wraptable}

Due to the extensiveness of the topic, not all papers could be included in this survey. Other notable works include, but are not limited to, the following references, \cite{dunlap2002revirt, ren2005distributed, vokorokos2006intrusion, vokorokos2010host}. 


\subsubsection{System Log IDS Research} 
Reuning~\cite{reuning2004applying} describes an anomaly detection system based on Bayesian probability theory and the term frequency inverse document frequency (TF-IDF) information retrieval technique. 
TF-IDF is applied to event log messages, treating each entry as an individual document. 
First, system training is required, in which data over a chosen time interval is collected and indexed into hash table, where each term is mapped to its TF-IDF weight. 
Messages with high scores, defined as the sum of the TF-IDF scores of the messages' terms, are detected. 
Results of the experiment on the DARPA99 dataset suggest that using log data solely produces a high false positive rate and many undetected attacks, but it can become a valuable component of a larger and more complex IDS.

Tchakoucht et al.~\cite{tchakoucht2015behavioral} improve upon the IDS of Yacine et al.~\cite{yacineEigen}.
The goal is to help decrease User-to-Root (U2R) and Remote-to-Local (R2L) attacks that exploit operating system or software vulnerabilities. 
User activity is audited based on LoginFlow, LoginFails, SessionDuration, SessionCPU, FormatCounter, AccessFails, DataVolume, and QuotaOverloadFails, which provide a feature vector representation for each user's behavior over a given time period. 
To characterize user behavior, $k$-means clustering identifies groups of similar user behavior. 
Euclidean distance is calculated to compare new user behavior to the reference profile in the detection phase. 
An experiment was conducted with a health information system consisting of three users, a patient, doctor, and an administrator, including their behavior over 30 days.  
The experiment resulted in significant improvements during learning and testing over Yacine et al.'s previous work \cite{yacineEigen} with a sizable increase in successfully identifying users and a large decline in false positives. 
Achieving good results, Tchakoucht et al. identify two constraints that can affect accuracy, change in user behavior and the system's inability to handle large datasets.

\subsubsection{System Log IDS Architecture Research} 
Guan et al.~\cite{guan2005collaborative} introduce KIT-I, an architecture for an IDS using system log data. 
Log data is stored in a secure server, so even if the computer is compromised, an intruder would not have the ability to modify log files to cover attack traces. 
The system consists of two modules, a transferring module and a neural network (NN) module. 
The transferring module is used to transfer log data at defined intervals from a client to a remote logging server via a secure channel, which is implemented using the SSL library in Java and a Certificate Authority for client to server authentication. 
The NN module is used to analyze received log files for abnormal behavior. 
In this work, no experiments are presented. 

Zhaojun \& Chao~\cite{zhaojun2010statistic} describe a new type of HIDS architecture based on an analysis of system logs. 
The architecture contains five modules\textemdash log collection and pre-processing, saving and updating, search and analysis, statistics and analysis, and alarming. 
During execution of the first four modules, system logs are collected and turned into records containing fields extracted from three parts of the system logs, namely,  a priority with ``Facility" and ``Severity" fields, a header with ``Timestamp" and ``Hostname" fields, and a message with ``Tag" and ``Content" fields. 
A constructed record is stored in a MySQL database and filtered using regular expressions to extract important records. 
For the decision engine, records from the database are 
transformed into numerical values and then passed through a back-propagation NN (BPNN) model for analysis. 
Once analysis is complete, the alarm module determines how to inform the user, if necessary.



Wang \& Zhu~\cite{wang2017centralized} propose a centralized HIDS architecture for private cloud computing, with the main goal to reduce usage of system resources. 
Their model is built on OpenStack\footnote{ See \url{www.openstack.org}.}, an open-source infrastructure platform for cloud computing) and consists of three nodes, compute, controller, and network nodes, and four modules, data collection, data pre-processing, detection, and alarm modules. 
The collection module uses Logstash\footnote{ See \url{https://www.elastic.co/products/logstash}.} to gather system logs from all VMs and stores it into Elasticsearch\footnote{ See \url{https://www.elastic.co/products/elasticsearch}.} for farther analysis by the detection center, which uses a C5.0 decision tree (DT). 
If an anomalous event is detected, the detection center alert to victim VM. 
This model was tested using the KDD99 dataset and compared to a traditional HIDS. 
Comparing the new centralized HIDS with a traditional HIDS shows that a centralized HIDS CPU utilization is approximately 14\% lower, memory consumption is about 2\% less, and the detection rate of 94\% is about the same with a slightly longer detection time.

\subsection{Audit Data IDSs}
In this survey, audit logs will refer to more granular information than system logs, collected with the goal of providing a chronological, detailed record of user activities.  
For example, audit logs allow visibility into network connections (e.g., source/destination bytes, protocols, etc.), command line actions (e.g., number of shells opened), privilege escalations, and changes to files. 
System calls are included in the audit logs, but the wealth of IDS research using them is discussed separately in the next section. 
Audit logs are high volume and costly to collect and manage, but they give higher fidelity for forensics and detection. 

\begin {table}
\small
\caption {HIDS with Audit Data} \label{tab:auditdata} 
\begin{tabular}{p{3.8cm} p{1.7cm} p{1.6cm} p{2.7cm} p{1.8cm} }
    \toprule

IDS Reference & Technique & Dataset & Classifier & Learning \\
  \midrule
Ilgun '93 \cite{ilgun1993USTAT} & Misuse & NA & Rules & NA \\
Ye et al. '01, '02 \cite{ye2001probabilistic, ye2002multivariate} & Misuse, Anomaly & DARPA98 + simulation & DT, $T^2$ test, $\chi^2$ test, Markov model & Both \\
Botha \& Von Solms '03 \cite{botha2003utilising} & Misuse & Self made & Rules + Fuzzy logic & NA \\ 
Li \& Manikopoulos '04 \cite{li2004windows} & Anomaly & Self made & OCSVM & Unsupervised \\
Shavlik \& Shavlik '04 \cite{shavlik2004selection} & Anomaly &  Self made & Winnow, NB & Supervised \\
Lin et al. '10 \cite{lin2010design} & Hybrid & NA & OSSEC, NN & Supervised \\
Mehnaz \& Bertino '17 \cite{mehnaz2017ghostbuster} & Anomaly & Self made & FSA rule-mining & Unsupervised\\

    \bottomrule
    
\end{tabular}
\end{table}


Ilgun~\cite{ilgun1993USTAT} illustrates a real-time IDS for UNIX operating system called USTAT (State Transition Analysis Tool for UNIX), which is the UNIX version of STAT described by Porras et al.~\cite{porras1992penetration}. 
It is a rule-based IDS and works by matching known patterns to the sequences of audit data gathered by the audit collection mechanisms of the OS. 
Some of the aims of USTAT are to automate a matching process and make patterns more flexible to adopt to different instances of equivalent attacks. 
This proposed IDS is able to detect attacks which involve cooperation of multiple user sections or accounts.
USTAT analysis is based on state changes, where state is ``the collection of all volatile, permanent, and semi-permanent data stores of the system at a specific time" and changes are called actions; therefore, an attack pattern is defined as a sequence of attacker actions. 
There are four main components, the data pre-processor, the knowledge-base component (containing fact-based data of objects of the system and rule-based data of state transitions), the inference engine (used to infer all states of the system and detect attacks), and a decision engine (used to chose an action and inform the user about results from inference engine).  
The conducted experiment was not focused on detection accuracy but instead on resources utilization running USTAT with other processes. 
This resulted in a limitation of disk throughput when running both USTAT and an audit daemon that collects audit trails.

Ye et al.~\cite{ye2001probabilistic} study data attributes for intrusion detection. Attributes include: (1) individual event occurrences (e.g., ``audit events, system calls, user commands''), frequencies (e.g., ``number of consecutive password failures''), and durations (e.g, ``CPU time of a command, duration of a connection''), (2) event combinations, (3) multiple events frequency and distribution, and (4) event sequence/transition.
They compare the intrusion detection performance of four methods\textemdash a supervised DT and three unsupervised anomaly detection algorithms utilizing both Hotelling's $T^2$ test ($T^2$ test) and the $\chi$-squared distance ($\chi^2$ test)\textemdash both multivariate statistical analysis methods, and a first order Markov model\textemdash
for intrusion detection in their experiments with the DARPA98 dataset and simulated attacks. The Markov chain based on an ordering property showed superior performance. This verifies that the ordering and frequency of audit events provides useful information to detect intrusions.

Follow-on work by Ye et al.~\cite{ye2002multivariate} present more results comparing $T^2$ test and the $\chi^2$ test, on audit trails to detect anomalous behavior. 
The proposed techniques are better in session-wise analysis (an entire session is considered an intrusion if it contains a single intrusive event), and overall performance of $\chi^2$ is better than $T^2$.

Botha \& Von Solms~\cite{botha2003utilising} implement a hybrid IDS based on comparing user actions with intrusion actions, using fuzzy logic.
These actions are interpreted as phases of an intrusion, which they describe using their own schema: "probing," "initial access," "super-user access," "hacking," "covering," and "backdoor."  These are represented as a graph for comparison using fuzzy logic. 
The authors developed a working prototype, and testing was done with the help of 12 users, where ten users were conducting both ``legal'' and ``illegal'' (presumably normal and unauthorized) activities and two users were conducting only legal activities. 
The system correctly identified both users conducting ``legal'' activities by assigning intrusion probabilities of 0\%, while the remaining users had probabilities of intrusive activities between 12\% and 48\%.

Li \& Manikopoulos~\cite{li2004windows} model user profiles with one-class support vector machines (OCSVMs), an unsupervised support vector machine (SVM) technique requiring only the user's own legitimate sessions to build  the user's profile, using a year of Windows audit data with a focus on masquerade detection. 
This approach allows for easier user management, such as adding and removing users legacy users, rather than multi-class classification methods. 
Results show the two-class training achieves a detection rate of 63\% with a 3.7\% false alarm rate and one-class training shows a 66.7\% detection rate with a 22\% false alarm rate. 
Even though the one-class training approach results in increased false alarms, this is offset by easier management and a reduction in training time.

Subsystems can monitor an abundance of system actions in the Windows OS.
An anomaly detection system is presented by Shavlik \& Shavlik~\cite{shavlik2004selection} that performs statistical profiling of users and system behavior on Windows 2000. 
Their algorithm uses measurements taken from two-hundred Windows 2000 attributes at one second intervals to generate approximately 1,500 features. 
Examples of features are encodings of CPU utilization, data input-output quantities, process information, and differences and averages of current versus historical values, among others. 
User behavior is accurately identified using features with assigned weights. 
Moreover, unique signatures are created by assigning individual user feature weights.

Winnow~\cite{littlestone1988learning}, a simple linear binary classification fitting algorithm, generally used with a large number of features, is tested against Na\"ive Bayes (NB) classifier. 
Succeeding training, all features ``vote" every second as ``for" or ``against" the likelihood of an intrusion occurrence. The weighted votes are then compared against a threshold to determine if there is an intrusion. 
During training, Winnow changes the weight of features that fired on incorrectly labeled instances, similar to perceptron training. 
Self-collected data from multiple hosts is used for gathering a baseline for normal users, and the same from a held-out set of hosts is labeled ``intrusions'', with the motivation of identifying insider threats. 
Winnow yields a 95\% detection rate with a low false-alarm rate (under one per day per computer), while NB has a 59.2\% detection rate and has 2 false alarms per day.


Lin et al.~\cite{lin2010design} propose an HIDS architecture combining open-source misuse detection with supervised learning. 
Misuse detection is implemented using OSSEC~\cite{OSSEC} (Open Source HIDS SECurity), an open-source IDS framework capable of conducting analysis of log files; anomaly detection is implemented using a BPNN. 
The misuse detection process consists of collecting log data, pre-processing and analyzing 
data with OSSEC, and finally reporting results. 
The anomaly detection process consists of training the BPNN. 
This can be trained with login or session activities, resource utilization, file operation activities, among other data types. 
BP training may take days or weeks and was cited as an area of active research. 
In this work, no testing was discussed. 



Mehnaz \& Bertino~\cite{mehnaz2017ghostbuster}  present Ghostbuster, a HIDS that profiles users based on their file-system access patterns and detects anomalies. 
The Linux utility \texttt{blktrace} \footnote{ See \url{https://linux.die.net/man/8/blktrace}.} is used to extract sequences of file access events. 
During the profile creation phase for each user, a feature vector is created by encoding file access by sizes (with blocks as units), frequencies, and patterns of files accessed. 
Statistical outliers of file access size and frequencies are a cause for alerts, and a finite state automata (FSA) for access patterns defines rule-based anomaly detection. 
Performance evaluation is given for actual file accesses of 77 users for 560 target files over eight weeks, four for training and four for testing. 
Results are given for many simulated attacks, and overall high detection rates and low false positive rates are reported; overhead is reported at 2\%.

We note that research for utilizing audit log data for intrusion detection in a cloud environment is a budding area of research, but is outside of the scope of this work \cite{mukkamala2002intrusion, lee2011multi, xu2016cloud, zhang2015cloudmonatt}. 

\section{System Call IDSs}
\label{sec:system-calls} 

System call data is a popular choice for HIDS research, because they are a primary artifact of the OS kernel; that is, there is no filtering, interpretation, or processing (such as, in the production of log files),  that can obfuscate events~\cite{creech2014semantic}.  
Often the unit of data used for detection is a system call trace,  a sequence of all calls invoked by a single process in a given time window. 
Hence, these IDS developments sit at an OS-level, but the object of modeling is program level. 
System calls can be collected for example with the `strace` utility, although there are many other ways to collect this same information.  
Some common calls include `open', `close', `read', `write', `wait', `exit', `mmap', among many others.  
Modern OSes often have hundreds of syscalls, for example the `syscalls' Linux manual page lists over 300.
Drawbacks of system-call-based approaches include the large computational overhead needed for harvesting and analysis and the large possible variations that potentially lead to false positives~\cite{apap2002detecting}.

Because each process produces a sequence of system calls, language modeling techniques are prevalent for system call-based HIDSs. 
In particular, many variations on $n$-gram features and Markov models of sequences of calls are configured to produce normal/attack classifications. 
See Forrest et al.~\cite{forrest2008evolution} for a more detailed survey of pre-2008 works leveraging system calls. 
Critical insights from this section are as follows: 
\begin{itemize}[leftmargin = *]
\item While research has shown that normal processes can be profiled using system calls, wide variations occur across processes, or for fixed processes across different user environments, installation configurations, etc. 
\item Detection results are quite sensitive to the length of sequence-based features with six- to eight-grams being strong choices.
\item  Short sequences provide less computation during training but are easier to bypass than longer sequences.  
\item Augmenting calls with other information, such as arguments of the calls, program counters, and addresses, can yield higher accuracy with less overhead. 
\end{itemize}
Overall, general trends indicate that features which model sequences are more costly than simple frequency counts of individual features, but yield better detection. 
Finally, meta-trends show that as labeled datasets become popular, a flurry of research ensues allowing IDS comparisons across papers and testing of many standard machine learning algorithms to flourish. 

\begin {table}[H]
\small
\caption {HIDS Leveraging System Calls} \label{tab:systemcalls} 
\begin{tabular}{p{4.5cm} p{1.4cm} p{1.5cm} p{3.3cm} p{1.8cm} }
    \toprule
IDS Reference & Technique & Dataset &  Classifier & Learning \\
   \midrule
Forest et al. `96 \cite{forrest1996sense}            & Anomaly           &   Simulated  &  Rules & Unsupervised \\
Kosoresow et al. `97 \cite{kosoresow1997intrusion}   & Anomaly    &   Self made        & FSA & Unsupervised \\
Hofmeyr et al. `98 \cite{hofmeyr1998intrusion}       & Anomaly    & Simulated + Self made & FSA & Unsupervised \\
Ghosh et al. `99 \cite{ghosh1999study, ghosh1999learning}& Hybrid & DARPA 99  &  NN & Both \\
Warrender et al. `99 \cite{warrender1999detecting} & Hybrid & UNM, DARPA98   & Rules, HMM & Unsupervised \\
Sekar et al. `01 \cite{sekar2001fast} & Anomaly & Simulated + Self made & FSA & Unsupervised \\
Wagner \& Dean `01 \cite{wagner2001intrusion} & NA & Self made &  Static Analysis & NA \\
Liao \& Vemuri `02 \cite{liao2002use} & Hybrid & DARPA98 &  $k$-NN & Both \\
Abad et al. `03 \cite{abad2003log} & Hybrid & Self made & RIPPER & Both \\
Feng et al. `03 \cite{feng2003anomaly} & Anomaly &  Simulated + Self made &  FSA & Unsupervised \\
Hoang et al. `03, `09 \cite{hoang2003multi} & Hybrid & UNM & HMM + Rules & Unsupervised  \\
Kruegel et al. `03 \cite{kruegel2003bayesian} & Anomaly & DARPA 99 & BN & Unsupervised \\
Kruegel et al. `03 \cite{kruegel2003detection} & Anomaly & DARPA 99 &  Probability models & Unsupervised \\
Jha et al. `04 \cite{jha2004filtering} & Anomaly & UNM & Filtering, Markov Models & Unsupervised\\
Tandon \& Chan `05, `06 \cite{tandon2005learning, tandon2006learning} & Anomaly & UNM, DARPA98 & Rules & Unsupervised \\
Han \& Cho `05 \cite{han2005evolutionary} & Anomaly & DARPA99 &  ENN & Unsupervised\\
Zhang et al. `05 \cite{zhang2005application} & Hybrid & DARPA98 & $k$-NN, Robust SVM, SVM, OCSVM & Both \\
Gao et al. `06 \cite{gao2006behavioral} & Anomaly & Self made & HMM-based distance & Unsupervised \\
Hu et al. `09 \cite{hu2009simple} & Anomaly & UNM, DARPA98 &  HMM & Unsupervised \\
Ahmed et al. `09 \cite{ahmed2009host} & Hybrid & UNM &  RBFNN & Supervised \\
Tong et al. `09 \cite{tong2009research} & Hybrid & DARPA & RBFNN + ENN & Supervised \\
Ye et al. `10 \cite{ye2010intrusion} & Anomaly &  NA &  Rules & Unsupervised \\
Jewell \& Beaver `11 \cite{jewell2011host} & Anomaly & Self made &  Rules & Unsupervised\\
Elgraini et al. `12 \cite{elgraini2012host} & Anomaly & UNM &  NB with a MM & Unsupervised \\
Xie et al. `13, `14 \cite{xie2013evaluating, xie2014evaluating-2, xie2014evaluating} & Anomaly & ADFA-LD &  $k$-NN, OCSVM, $k$-Means & Unsupervised \\
Creech \& Hu `14 \cite{creech2014semantic} & Anomaly &  UNM, ADFA-LD, KDD98 & ELM NN & Supervised \\
Anandapriya \& Lakshmanan `15 \cite{anandapriya2015anomaly} & Anomaly & ADFA-LD & SVM, ELM NN & Supervised \\
Gupta \& Kumar `15 \cite{gupta2015immediate} & Misuse & UNM & Rules  & Unsupervised \\
Haider et al. `15 \cite{haider2015towards} & Anomaly & ADFA-LD &  $k$-NN & Unsupervised \\
    \bottomrule
\vspace{-1cm}
\end{tabular}
\end{table}

\subsection{Sequential Features (\textit{n}-Grams)}
Early work of Forrest and Longstaff~\cite{forrest1996sense} provide preliminary HIDS results by characterizing normal (frequent) and then identifying abnormal (infrequent) short sequences of System calls. 
One way to conceptualize the main idea is that System calls are ``words'', sequences of calls form ``phrases''. 
The general trend incurs relatively large computational expense for feature extraction and/or model training, but reap strong detection metrics.

Similar anomaly detectors based on modeling $n$-grams of system calls were explored by others, but without statistically modeling their frequencies~\cite{hofmeyr1998intrusion, kosoresow1997intrusion}. 
Helman \& Bhangoo~\cite{helman1997statistically} rank system call traces by the likelihood of $n$-grams in normal versus attack scenarios. 
Ye et al.~\cite{ye2010intrusion} use set theory to design an algorithm that learns rules defining normal system call sequences, then detect anomalies based on votes from the rules, although no testing is presented. 

For each process, Jewell \& Beaver~\cite{jewell2011host} consider variable length sequence of System calls, defined as an observed sequence of System calls for which no call occurs twice. 
Comparing this with other sequential features, (e.g., $n$-grams), they observe that the counts of the system call sequences observed plateau for normal user activity faster than other definitions, and  the counts spike upon novel activity. 
With the goal of identifying malicious data exfiltration activities in real-time, an experiment in which researchers were challenged to exfiltrate three file collections on a given set of machines over two days is used to collect malicious and normal system call data, which is used to validate the approach. 


Elgraini et al.~\cite{elgraini2012host} estimate the probability of a sequence of calls conditioned on the class (normal/attack) using a first order Markov model\textemdash $P(s_1, s_2, ... | C) = P(s_1|C)P(s_2|s_1, C)... $. Finally, a NB approach is used to find the most likely class. Results are compared to many other previous classifiers on data from the University of New Mexico (UNM), finding that this method performs similarly.

Creech \& Hu~\cite{creech2014semantic} make two innovations for a HIDS based on kernel level system call traces, (1) creating semantic features of system call sequences (phrases) by defining a context free grammar and (2) using an extreme learning machine (ELM)\textemdash a neural network (NN) classifier of Huang et al.~\cite{huang2004extreme}. 
This approach takes per-host training that is computationally costly, taking days or weeks, although once trained, labeling (or decoding) is fast and accuracy results are very strong, reported via the Receiver Operating Characteristic (ROC) curve using the Darpa98\footnote{Darpa98 is sometimes referred to in other literature as KDD98.} and ADFA-LD datasets. 
Anandapriya \& Lakshmanan~\cite{anandapriya2015anomaly} also test anomaly detection results using semantic features with the ELM on the ADFA-LD dataset.

Gupta \& Kumar~\cite{gupta2015immediate} define a signature for a program as the admissible bigrams of calls, specifically those seen in training. This allows lightweight detection of programs with a variety of new two-sequences of calls that gives highly accurate results as tested on the UNM dataset. Their work discusses implementation for cloud infrastructure using multiple VMs.

\subsection{Frequency-Based Features: A Cheaper Alternative} 
In response to the costly but effective sequence-based features, research to develop and test more computationally inexpensive, frequency-based features from system call traces finds, at least for the AFDS-LD dataset, that such features still produce strong accuracy results. 

Liao \& Vemuri~\cite{liao2002use} regard traces as documents represented with the vector of TF-IDF scores for each word (system call). 
The $k-$nearest neighbor ($k$-NN) with cosine similarity distance is used for anomaly detection. 
If a process is classified as intrusive, the whole session it belongs to is also considered an attack session.
Liao et. al performed the experiment using the names of System calls recorded in Basic Security Module\footnote{ See \url{https://docs.oracle.com/cd/E19457-01/801-6636/801-6636.pdf}.} (BSM) audit data from DARPA98 dataset; 
they exhibited over 90\% TPR with under 2\% FPR. 
The second experiment preempted this TF-IDF anomaly detector with signature verification. 
First, each process is compared to a set of abnormal processes using cosine similarity, and, if they match, the process is marked as intrusive. 
Otherwise, the $k$-NN anomaly detection process is used to classify the process.
This two-stage workflow produced 91.7\% detection rate and 0.59\% false positive rate with threshold of 0.8.
This method is computationally efficient, with complexity $\mathcal{O}(N)$, with $N$ as the number of processes.


Continuing the system call/trace interpretation as words or documents, respectively, Zhang et al.~\cite{zhang2005application} propose two novel techniques to lower false positives. 
First, a modified TF-IDF score is crafted from system call traces; 
second, the authors build a detector using supervised training with Robust SVMs to battle noisy training data, OCSVMs for unsupervised training, and $k$-NN. 
Online-training of the SVMs is used to decrease training time while preserving accuracy of intrusion detection. 
Clean and noisy datasets are generated from system calls of privileged processes in the DARPA98 dataset, these are used to compare each classifier with and without their modifications. 
Results showed that the detection accuracy of the modified classifiers is the same or higher than the baseline, when tested with both the clean and noisy datasets, while the training time ratio for the modified SVM over the original is between 51.61\% and 66.67\% (i.e., retraining is significantly faster). 

Xie \& Hu, and Xie et al.~\cite{xie2013evaluating, xie2014evaluating-2, xie2014evaluating} consider simple features such as a trace's length, and the relative frequency of each call in that trace, and achieve ``acceptable'' detection results (i.e., ROC curves) in their testing with the ADFA-LD dataset, using simple one-class classifiers; namely, $k$-NN, OCSVM, and $k$-means algorithms. 

Haider et al.~\cite{haider2015towards} propose using different, but still inexpensive, statistical features on system call traces of the ADFA-LD dataset, with the same goal of fast performance of transforming data to features without sacrificing accuracy of detection. 
Four features, namely, the least/most repeated and the minimum/maximum values in a trace, are used to represent a trace to detect attacks, and three supervised learning algorithms, SVM with linear and radial basis kernels and $k$-NN, are used. Results show $k$-NN receives a 78\% TPR, average(FPR, FNR) = 21\%. 
These results increased the TPR over works of Xie \& Hu, and Xie et al.~\cite{xie2013evaluating, xie2014evaluating-2, xie2014evaluating}, for similar false positive metrics, but are far less accurate than the computationally expensive work of Creech et al. \cite{creech2014semantic}. 
Although this set of work used the same dataset, it is not clear from the authors' treatment if the experiments provide a fair comparison across papers.

\subsection{Hidden Markov Models (HMMs) for System Call Modeling} 
HMMs are a natural data model for sequential data and many other works employ HMMs for system-call-based IDSs. 
Warrender et al.~\cite{warrender1999detecting} compare four methods of detection based on $n$-grams of system call traces: list-and-lookup of observed sequences, relative frequencies, RIPPER rule induction algorithm of Cohen~\cite{cohen1995fast}, and HMMs. 
Their conclusions indicate that sufficiently accurate detection results are achievable by more computationally efficient algorithms than HMMs, and that accuracy results are more dependent on test datasets than the algorithm chosen. 

Gao et al.~\cite{gao2006behavioral} create a novel HMM-based metric that reports better IDS results than their previous ``evolutionary distance (ED)'' metric, while also obtaining 6\% faster performance. 

Hoang et al.~\cite{hoang2009program} develop a hybrid detection scheme that uses both a HMM to model system call sequences and a ``normal'' database, which includes the frequency of each observed database short sequence. 
Fuzzy rules are defined to classify a newly-observed sequence and take into account the sequence's probability (computed via the HMM) and frequency in the normal database. 

HMM training is performed using an incremental method in conjunction with an initial parameter optimization method
to reduce the high cost incurred during computation.  
Validation on the AFDA-LD dataset exhibit a lowering of false positive rate of 48\% while indicating greater anomaly detection than a ``normal-sequence database scheme and a two-layer scheme.'' 
In addition, the HMM training time realized a 75\% reduction while simultaneously decreasing the memory usage. 
This hybrid approach follows their earlier work \cite{hoang2003multi}, where first the ``normal'' database is used to determine frequency, and second HMM-likelihood is computed detect anomalies of only those sequences of system calls that are rare or unseen in training. 
Experiments on the UNM's dataset (only using sendmail program traces) prove that this approach is better in detecting anomalous behavior of programs in terms of accuracy and response time than a conventional single layer approach; however, the HMM model training is expensive. The integrated system is able to produce higher levels of anomaly signals as soon as an intrusion occurs. Known problems include storage requirements, reducing the training cost of the HMM, and determining the parameters of the model automatically. 

Hu et al.~\cite{hu2009simple} propose a pre-processing and training approach for HMMs that halves training time of traditional HMMs with ``reasonable'' accuracy, i.e., with some adverse effect to the false detection rate as tested on UNM and DARPA98 datasets. 
In general, the work breaks training sequences into many small sequences, and train many ``small'' HMMs, and finally take take a weighted average.

\subsection{Other System Call IDS Works}
Jha et. al.~\cite{jha2004filtering} introduce a novel statistic-based anomaly detection algorithm for system call sequences. 
They observe that after Markov models (not HMMs)  are learned from observed sequences of System calls, an observed sequence is assumed to be a mixture of the learned models and a chaotic model. 
Bayesian techniques are used to optimized the mixing parameter, and if it is greater than a specified threshold, an alert is raised. 
By using mixtures of Markov chains, their filtering approach can model mixtures of system call traces from multiple users, potentially in cases involving multiple users cooperating.  
Additionally, the filtering based approach can address the masquerade-detection problem, allowing for the identification of the user that generated a given execution trace based on usage patterns. 
Results for many configuration parameters are given on the UNM data set. 
Comparing this technique to HMMs, one finds that Markov chain training  is $O(m)$, with $m$ the length of the trace, while HMM training has complexity $O(n*m^2)$, where $n$ denotes the number of HMM states. 

Ghosh et al.~\cite{ghosh1999study, ghosh1999learning} test artificial neural networks (ANNs) for misuse (supervised) and anomaly detection (unsupervised) using the DARPA99 dataset. 
For anomaly detection, a NN is trained using normal data and randomly generated data (for simulated attacks). 
ROC curves are given showing strong results, notably, a TPR of 77.3\% and a FPR of 2.2\%. 

Han \& Cho~\cite{han2005evolutionary} introduce an IDS utilizing evolutionary neural networks (ENNs) to simultaneously calculate the NN’s structure and weights.  
For labeled training data, ambient system call sequences are labeled normal (non-attack) and randomly generated sequences are labeled anomalous (attack) at a rate of 2-to-1. 
Experiments with an ENN produced a 0.0011\% false-alarm rate while obtaining a 100\% detection rate using the DARPA99 data set.
Performance shows that training the ENNs takes about an hour; in comparison, this is about order of magnitude longer than training any single, comparably structured NN, but about an order of magnitude less than a grid search over many traditional NNs.

Tong et al.~\cite{tong2009research} propose a new hybrid IDS using Radial Basis Function (RBF) NN with Elman NNs (ElNN) for both anomaly and misuse detection. 
RBFNN classifies events in real time, passing output as an input into the ElNN. 
Positively (respectively, negatively) detected events by the RBFNN increase (respectively, decrease) a context weight in the ElNN, which improves accuracy and decreases the false positive rate. 
This technique is advantageous due to its memory of prior seen sequences\textemdash it is robust to sparse occurrences of misuse or anomalies but will detect high temporal density of anomalies and misuse\textemdash  and it exhibits faster training time, as compared to the Multilayer Perception (MLP) NN IDS of Ghosh et al.~\cite{ghosh1999study}. 
Evaluations with the DARPA dataset resulted in an anomaly detection accuracy of 93\%, false positive rate of 2.6\%, and a misuse detection accuracy of 95.3\% with a 1.4\% false positive rate. Results were compared to \cite{ghosh1999study} and \cite{jirapummin2002hybrid}, ultimately producing higher accuracy and lower false positive rates.

Ahmed \& Masood~\cite{ahmed2009host} test radial basis function NNs on the UNM dataset, exhibiting accurate detection. 
Explicitly, they optimize $y(x) = \sum_1^N w_i \phi_{\sigma_i} ( x - \mu_i )$, for a spherical radial basis function $\phi$ centered at $\mu_i$ with variance $\sigma_i^2$ (i.e., $\phi_{\sigma_i}(x) = \exp(-\sigma_i^{-2} \|x\|^2) $) to learn $w_i, \sigma_i, \mu_i$, and they augment the training algorithm to also learn $N$, the number of basis functions.

Wagner \& Dean~\cite{wagner2001intrusion} use static analysis to automatically derive three models of application's system call behavior. 
Immediate detection of a program's wrongful behavior allows for the detection of intrusions. 
More generally static analysis is a large area of research that is outside the scope of this HIDS survey. See other static analysis surveys~\cite{idika2007survey, moser2007limits}.

Kruegel et al.~\cite{kruegel2003bayesian} create anomaly detectors modeling four features of system calls. These four detectors outputs are dependent nodes in novel Bayesian network (BN), along with dependent nodes for the four detectors confidence, and a single independent node for the classification. 
Results show perfect detection rates with a 0.2\% FPR. 
Training time is costly (NP-hard), but labeling is $\mathcal{O}(N)$ with $N$ as the number of nodes in the network.

\subsection{Using System Call Arguments and Additional Data} 
In addition to modeling the system calls, incorporating the arguments of the calls or memory pointers has garnered IDS results. 

Abad et al.~\cite{abad2003log} describe an IDS based on correlating network traffic to system calls and aiming to increase the detection rate and decrease the false positive rate for both misuse and anomaly detection. Two approaches were taken, top-down, where attacks' behavior is analyzed to identify which logs can contain evidence of attack, and bottom-up, where multiple logs are analyzed to detect a specific attack. The bottom-up approach finds attacks through log correlation, and since logs may have millions of entries, the RIPPER data mining tool is used for record filtration.
To conduct the experiment, the authors used RIPPER, a rule mining algorithm that attempts to ``predict the next system call", combined with log correlation using both System calls and network traffic.
These ideas follow from  work of Lee \& Stolfo \cite{lee1999data}. 
Results show an increased detection rate and a decreased false positive rate.

For each system call (e.g., read, write, ..) for each process (e.g., sendmail), Kruegel et al.~\cite{kruegel2003detection} build models of normal arguments' string lengths, characters, and structure. 
Similar features found in Kruegel et al.~\cite{kruegel2003bayesian} are used on System calls, not arguments.   
For anomaly detection, arguments with sufficiently different features are flagged, and the detector exhibits strong detection accuracy. 
Overhead is investigated, showing about 5Kb of memory is required, and 18\% (of a 2003 era) processor was used. 
Follow-on research of Mutz et al.~\cite{mutz2006anomalous} uses the same Bayesian network of Kruegel et al.~\cite{kruegel2003bayesian} to combine these system call argument feature anomaly detectors into an ensemble. 

Tandon \& Chan~\cite{tandon2005learning, tandon2006learning} develop an anomaly detection system based on rule learning techniques that leverage both system calls and their arguments. 
Results show gains over using just System calls, but at significant computational expense (an order of magnitude higher). 
Similarly, other works leveraging the arguments of System calls to enhance system-call-based detectors became prevalent at this time; e.g., see Bhatkar et al.~\cite{bhatkar2006dataflow} and Sufatrio \& Yap~\cite{sufatrio2005improving}. 

Sekar et al.~\cite{sekar2001fast} use finite state automata (FSA) to model the programs' code path by combining System calls (transitions between states) with program counter information (to learn states). 
This is a computationally cheaper approach than the HMM and $n$-gram techniques, and also improves accuracy over these techniques. 
To create a model of the virtual path between calls, Feng et al.~\cite{feng2003anomaly} incorporate dynamic extraction of return addresses in addition to the FSA approach, yielding additional accuracy without increased cost.

\subsection{System Call Mimicry Attacks} 
Finally, we note that system-call IDS developments are met with research designing attacks to evade such measures, with key ideas including ``mimicry'' attacks, where null-effect calls pad the malicious sequence of effective calls~\cite{giffin2006automated, kayacik2006evolving, kayacik2007contribution, kayacik2008mimicry, kruegel2005automating, wagner2001intrusion} or malicious call sequences are sufficiently small to evade detection~\cite{tan2002why, tan2002undermining}.

 
\section{Windows Registry IDSs}
\label{sec:registry}

Windows Registry is the OS's key-value database containing configuration settings for all programs and hardware on that host. 
This database is heavily used during computer operation. 

All processes use the Registry, including malware that also often modify the Registry to achieve their aim~\cite{hoglund2006rootkits}. 
Consequently,  Registry monitoring has been leveraged by many researcher efforts for forensic analysis~\cite{carvey2005windows, dolan2008forensic, mee2006windows}. 
Below we survey the few works that build HIDS from Registry data.

Initial work by Apap et al.~\cite{apap2002detecting} proposed the Registry Anomaly Detection (RAD) system, consisting of three components, an audit sensor to log Registry activities, a model of normal behavior, and a real-time anomaly detector. 
RAD extracts five raw features from Registry accesses, namely: (1) the process accessing the registry, (2) the query type requested, (3) the key used, (4) its value, and (5) the outcome (e.g., success, error) called the response.
Importantly, any anomaly detection algorithm that can accommodate these sparse feature vectors is applicable.
In this initial work, the probability of each feature (5 distributions), and conditional probability of pairs of features (20 distributions) are estimated following Friedman and Singer~\cite{friedman1999efficient}, and the detection system alerts if any of the 25 estimates are below a threshold. 
An advantage of this estimation is that models are continually updated without any user interaction.

Heller et al.~\cite{heller2003oneclass} and a follow-on publication of Stolfo et al.~\cite{stolfo2005comparative} both test OCSVMs for the anomaly detection component of RAD with three different kernels and conclude that the probabilistic anomaly detector (PAD) of Apap et al. is much more accurate. 
Computational analysis is also given. 

\begin{table}[!ht]
\begin{center}
\small
\caption {Registry Anomaly Detection Systems} 
\label{tab:reg} 
\centering
    \begin{tabular}{p{3.1cm} p{1.4cm} p{1.8cm} p{1.5cm}}
    \toprule
    IDS Reference & Technique & Performance Cost & Memory Cost  \\
   \midrule
     Apap et al. `02 \cite{apap2002detecting} & PAD & $\mathcal{O}(v^2d^2)$ & $\mathcal{O}(vd^2)$\\
     Heller et al. `03 \cite{heller2003oneclass} & OCSVMs & $\mathcal{O}(dL^3)$ &  $\mathcal{O}(d(L+T))$ \\  
     Stolfo et al. `05 \cite{stolfo2005comparative} & OCSVMs & $\mathcal{O}(dL^3)$ &  $\mathcal{O}(d(L+T))$ \\ 
     Topallar et al. `04 \cite{topallar2004host} & SOM & & \\ 
    \bottomrule
    \end{tabular}
    \end{center}
\begin{tablenotes}
    \item \small \noindent Here $v$ denotes number of unique records,  $d$ the number of features, $L$ the number of training records, and $T$ the number of testing records. 
\end{tablenotes}
\end{table}

PAD takes time $\mathcal{O}(v^2d^2)$ and space $\mathcal{O}(vd^2)$ where $v$ denotes the number of unique records, and $d$ denotes number of record components (dimension). The OCSVM takes time $\mathcal{O}(dL^3)$ and space $\mathcal{O}(d(L+T))$ where $ L, T$ denote the number of training records, and  testing records, respectively. The comparison of algorithms was conducted on Pentium Celeron with 512MB RAM with memory usage of under 3MB, and 3\%-5\% of CPU usage.

Topallar et al.~\cite{topallar2004host} refer to the RAD system, but propose the use of Self-Organizing Maps (SOM), a NN model, as an algorithm for anomaly detection. The abstract claims their results demonstrate a low false positive rate in comparison to other IDSs.

\section{File System IDSs}
\label{sec:filesystem}
This section surveys work that propose or test IDSs that monitor host file-systems for detection. 
File systems have visibility to stored data, executables, and metadata used to service file requests. 
Malicious actions often involve modifying or adding new files or metadata (e.g. to allow unauthorized future access or remove evidence of previous access), leveraging file systems to monitor files, access to files, or determine legitimacy of any requests to the file system is a promising avenue for intrusion detection and prevention. 
Since file-system IDSs are logically separate from the OS, they are harder to disable and allow monitoring after compromise. 
The primary drawback of storage-based IDSs is their limited visibility.  
\begin{table}
\begin{center}
\small
\caption {File System Detection Systems} 
\label{tab:file-system} 
    \begin{tabular}{lll}
    \toprule
    IDS Reference & Name & Rule Base   \\
   \midrule
     Kim \& Spafford `94 \cite{kim1994design}           & Tripwire              & Checksum\\
     Griffin et al.`03 \cite{griffin2003feasibility}   & Disk-Based IDS   & Policy \\ 
     Pennington et al. `03 \cite{pennington2003storage} & Storage-Based IDS & Checksum \& policy\\
     Patil et al. `04 \cite{patil2004i3fs}              &    I\textsuperscript{3}FS & Checksum \& policy\\
    \bottomrule
    \end{tabular}
    \end{center}
\end{table}

First available in 1992, Tripwire\footnote{ See \url{www.tripwire.org}.}, from Kim \& Spafford~\cite{kim1994design},  is perhaps the most notable file-integrity tool. 
Tripwire is an open-source and now commercially available IDS for detection and remediation of malicious file and configuration changes originally designed for the UNIX system. 
A checklist of information about important files is created periodically and compared against previous versions to detect unexpected or unauthorized file changes. 
Details of the original system implementation and use are reported in the publication cited above. 
Notably, the system was deployed and in use before the publication.

Griffin et al.~\cite{griffin2003feasibility} implement ``IDS functionality in the firmware of workstations' locally attached disks," where the majority of system files lie. 
The Intrusion Detection for Disks (IDD) system monitors the file system for suspicious file manipulations, such as unauthorized reads, writes, file meta-data modifications, suspicious access patterns, compromises of file integrity, or other events which may indicate an intrusion. 
Since this IDS is required to run on separate hardware, it is protected even if the system it is monitoring has been compromised, so long as the storage device and administrative computer are uncompromised. 
The system has four main design requirements: specifying access policies, securely administrating the IDD, monitoring, and responding to policy violations. The system's architecture consists of three main components: (1) the bridge process on the host computer, to connect the administrator and IDD, (2) the request de-multiplexer, to differentiate administrative requests from other requests, and (3) a policy manager on the IDD, to monitor the system for violations and generate alerts.
An evaluation using a prototype disk-based IDS into a SCSI (Small Computer System Interface) disk emulator and using PostMark trans and SSH-build filesystem benchmarks indicates that it is feasible to include IDS functionality in low-cost desktop disk drives, in terms of CPU and memory costs. 

Pennington et al.~\cite{pennington2003storage} propose an Intrusion Detection on Disk, a rule-based IDS embedded in the storage interface and monitoring the file system.  
The system prototype uses a set of rules to monitor important files and binary changes (following Tripwire~\cite{kim1994design}) and rules to detect patterns of changes to the file system. 
Testing on 16 rootkits and two worms shows that 15 are identified by the IDS, and three of the detected 15 modify the kernel to hide from other file-system integrity checkers (e.g., Tripwire). 
Examples of alerting activities include ``modifying system binaries, adding files to system directories, scrubbing the audit log, or using suspicious file names.''  
The overhead of the system is investigated, and results show under 2MB of memory is needed. 
The primary advantages of this storage-based IDS are its independence from the host (if the host system is compromised, extra steps are necessary to disable this IDS), and that modifications to the storage device are necessary if any malware is to persist across reboots.

Patil et al.~\cite{patil2004i3fs} describe I\textsuperscript{3}FS, an In-kernel Integrity checker and Intrusion detection File System; this is an IDS based on real-time, in-kernel, on-access integrity file checking. 
The proposed IDS is modular and can be mounted on any file system. 
The main goal is to restrict access and notify administrators if an intrusion is detected. 
The system is compared against Tripwire~\cite{kim1994design} and can overcome its limitations\textemdash intruder tampering, large performance overhead, and inability of real-time detection. 
I\textsuperscript{3}FS uses security policies and cryptographic checksums of files computed using MD5, and stores both in four in-kernel Berkeley databases: policy, checksums, checksum metadata, and access counter databases. 
IDS security is implemented by adding an authentication mechanism which allows for file calls interception and by using policies and previously computed checksums to determine file integrity to allow or deny access to those files, and possibly alerting system administrators. I\textsuperscript{3}FS is primarily designed to prevent replacement of legitimate files with files containing malicious content, unauthorized modification of data, and data corruption. The system was tested using CPU, I/O, and custom read benchmarks. Results indicate that performance overhead under normal user workload is 4\%, and can be modified by setting system parameters and changing system policies.


File system monitoring is frequently used in HIDSs that leverage virtual environments. 
Quynh \& Takefuji~\cite{quynh2007novel} propose monitoring a system by implementing   sniffing and forwarding file system call logs (e.g., map, open, write)  to a privileged VM. 
Ko et al.~\cite{ko2011flogger} design a ``file-centric logger'' that watches file accesses and transfers and can be implemented in cloud VMs and physical environments.  
A tool is provided for the end user to verify personal file tampering. 
Gupta et al.~\cite{gupta2012secure, gupta2012light} describe a lightweight and platform independent HIDS based on monitoring file system integrity while running as privileged VM.
Jin et al.~\cite{jin2013vmm} implement VMFence, which includes file integrity monitoring, among other

(network-oriented) features.

Distributed and more comprehensive IDS architectures leveraging file-integrity for detection exist as well, see Demara et al. \cite{demara2004mitigation}. Their work also provides a short survey of existing frameworks for file-system IDSs.
\section{Program Analysis and Monitoring Techniques}
\label{sec:program} 
This section focuses on a few works that that leverage information about processes, process trees, or specific binaries on a host for detection. 
We note that this has significant overlap with other security sub-fields, such as dynamic malware analysis and application vulnerability analysis. 
A detailed survey of these related topics is out of scope for this survey. 

\begin{table}[ht]
\begin{center}
\small
\caption {Program Analysis for Detection Works} 
\label{tab:program-analysis} 
    \begin{tabular}{llll}
    \toprule
    IDS Reference & Data Source & Classifier & Learning\\
   \midrule
     Schultz et al. `01 \cite{schultz2001data}  &  Binary, DLL, calls  & RIPPER, NB, MNB & Supervised\\
     Newsome \& Song `05 \cite{newsome2005dynamic}  &  Binary  & Rules & Unsupervised \\ 
     Moscovitch et al. 07 \cite{moskovitch2007host} & Program's resource utilization & DT, NB, BN, ANN & Supervised\\
     Khan et al. `16, '17 \cite{khan2016fractal, khan2017cognitive}    &  Process' network utilization  & AdaBoost & Supervised\\
     Vaas \& Happa `17 \cite{vaas2017detecting} & & \\
    \bottomrule
    \end{tabular}
    \end{center}
\end{table}

\begin{wraptable}[13]{r}{0.53\linewidth}
\vspace{-.3cm}
\centering
\small
\caption {Schultz et. al. \cite{schultz2001data} Results} \label{tab:schultz-results} 
\begin{tabular}{ p{1.3cm} p{2.7cm} p{1cm} p{1cm}}
    \toprule
      Algorithm & Feature Type & TPR & FPR \\
    \midrule
    Signatures & Bytes & 33.75\% & 0\% \\
    RIPPER & DLLs & 57.89\% & 9.22\% \\
    RIPPER & DLLfunction calls & 71.05\% & 7.77\% \\
    RIPPER & DLLs with counted function calls & 52.63\% & 5.34\% \\
    NB & Strings & 97.43\% & 3.80\% \\
    MNB & Bytes & 97.76\% & 6.01\% \\
    \bottomrule
\end{tabular}
\begin{tablenotes}
    \item \small \noindent Here TPR is true positive rate, and FPR is false positive rate. 
\end{tablenotes}
\end{wraptable} 

Schultz et al. \cite{schultz2001data} describe a framework for automatic detection of malicious executables before they run.  
Different data mining algorithms are explored to determine the best algorithm for new binaries.

Experiments used three data mining algorithms; RIPPER, NB, and Multi-Na\"ive Bayes (MNB), and five types of features\textemdash Dynamically Loaded Libraries (DLL) used by the binary, DLL function calls made by the binary, number of unique function calls within each DLL, strings extracted from binary files, and byte sequences. 
To conduct the experiment, a dataset of malicious and benign executables were created from McAfee's virus scanner. 
Results were compared to conventional signature based detectors and are summarized in Table~\ref{tab:schultz-results}.

TaintCheck, a system that can ``perform dynamic taint analysis by performing binary rewriting at run time'' was developed by Newsome \& Song~\cite{newsome2005dynamic}. 
Data originating from or influenced by any input, e.g., memory addresses and format strings that are not supplied by the code itself, i.e., are supplied by external inputs or mathematical computation, are considered tainted, and when used unsafely indicates likely vulnerable code. 
TaintCheck identifies tainted code, then monitors instructions that manipulate it (e.g, MOVE, LOAD, PUSH instructions), and finally identifies if the data is used in a manner that violates set policies (e.g., as input to a system call). 
It reliably detects most types of exploits while producing no false positives, and permits  semantic analysis using signatures.

Moscovitch et al.~\cite{moskovitch2007host} test four machine learning techniques: DT, NB, BN, and ANN. 
Each has different feature subsets to detect unknown malware based on characteristics of known malware, in particular worms, using computer measurements, such as memory usage, disk usage, CPU usage, etc. 
To examine worm behavior, the authors used five known worms,
which all perform port scanning and other actions. 
A variety of configurations were created, using machines with different hardware and using different levels of activity from background tasks and user tasks.
Four hypotheses were tested: 
 the method can reach detection accuracy of known malware above 90\% and detection accuracy of unknown worms above 80\%;
the computer configuration and background activities have no significant influence on detection; 
furthermore, at most 30 features are needed to attain the same accuracy as full set. All goals were achieved, with BNs consistently producing accurate results.

Khan et al. \cite{khan2016fractal} introduce an anomaly detection HIDS based on a modified AdaBoost ensemble classifier. 
They add an ``information fractal (cognitive) dimension approach'', which  assigns higher weights to weak classifiers, which puts emphasis on misclassified samples to improve their estimation. 
This idea stems from fractal dimension theory~\cite{kinsner2005unified}.
A host sensor is utilized to collect the network profile of processes (process ID, time started, and the process's network connection information) and modules on Windows 7 OS.

To conduct the experiment, authors collected 333,692 data samples for 1 hour and used malware detected by 3 out of 54 antivirus companies, according to VirusTotal\footnote{ See \url{www.virustotal.com}.}. 
To build a classifier, the dataset was partitioned into 70\% and 30\% as training and testing sets, respectively. 
Results indicate that the proposed AdaBoost algorithm reduces the error 60\% more than the traditional algorithm with 30 less iterations. 
A comparison shows an improvement in detecting true positives from 93.93\% to 95.27\% and a reduction of false negatives from 6.06\% to 4.73\%; however, detection of true negatives decreased from 100\% to 97.14\% and false positives increased from 0.0\% to 13.7\%
A similar fractal approach by many of the same authors, Siddiqui et al.~\cite{siddiqui2016detecting} uses $k$-NN with a fractal weighting approach on network data for detection, and further work in Khan et al.~\cite{khan2017cognitive} attempting to model polymorphic malware with fractal analysis of the process tree.

Vaas \& Happa~\cite{vaas2017detecting} design a client-server architecture that observes process' memory consumption. 
Snapshots of each process are collected over a time window containing its resource utilization and timestamp, with the goal of identifying anomalous behavior of a machine's processes on a per-application basis. 
The method consists of three phases: acquisition, learning, and production. 
A memory fingerprint is gathered during the acquisition phase.
During the learning phase, a model of each application is computed from the fingerprint 
and the model is used to create an anomaly detector. 
During the production phase, the quality of the model is assessed.
The model is then tested with user process data, and results indicate an ability to distinguish processes by their virtual memory fingerprints. 
To increase efficiency during the learning phase, in order to make application models available more quickly, parallel machine learning techniques are utilized.

\section{Host-Relevant Algorithms Tested on Network Traffic}
\label{sec:other-data}
This subsection surveys select works that focus on algorithmic development. 
Most focus on advancing detection metrics or performance as tested on the DARPA- and KDD-related datasets; hence, their presentation falls under NIDS applications, but the detection algorithms can be applied to HIDS. 
Note a progression in the literature from testing standard machine learning algorithms against each other, to testing ensembles to gain accuracy, to parallelization and GPU acceleration efforts, and finally to pre-composing the classifiers with data-driven feature selection algorithms. 
For the interested reader, we note that this section complements the survey of Buczac \& Guven~\cite{buczak2016survey}, which also surveys only a minority of the many works focusing on machine learning advancements for detection, most of which are also tested on NIDS datasets, though Buczac \& Guven focus on more detailed analyses of the algorithms involved. 

\begin {table}
\small
\caption {Selected algorithmic developments for IDS, not tested on host data but  applicable} \label{tab:NIDS} 
\begin{tabular}{ p{3.8cm} p{1.4cm} p{3cm} p{2.5cm} p{2cm} }
    \toprule
      IDS Reference & Technique & Dataset &  Classifier & Learning \\
    \midrule

Barbar\'a et al. '01 \cite{barbara2001adam} & Anomaly & DARPA99 & Rule mining, DT & Supervised \\
Ambwani '03 \cite{ambwani2003multi} & Hybrid & DARPA98 & SVM & Supervised \\
Amor et al. '04 \cite{amor2004naive} & Hybrid & KDD99  & NB & Supervised \\
Liu et al. '04 \cite{liu2004genetic} & Anomaly & KDD99 & Genetic Clustering & Unsupervised \\
Chen et al. '07 \cite{chen2007hybrid} & Misuse & KDD99, DARPA98 &  NN 
& Supervised \\
Kayacik et al. '07 \cite{kayacik2007hierarchical} & Anomaly & KDD99 &  Hierarchical SOM & Unsupervised \\
\"{O}zyer et al. '07 \cite{ozyer2007intrusion} & Hybrid & KDD99 & Rules+boosting &  Supervised \\
Peddabachigari et al. '07 \cite{peddabachigari2007modeling} & Anomaly & KDD99 &  DT, SVM, DT-SVM, Ensemble & Supervised  \\
Tajbakhsh et al. '09 \cite{tajbakhsh2009intrusion} & Hybrid & KDD99 &  ABC & Supervised \\
Wang et al. '10 \cite{wang2010new} & Hybrid & KDD99 & FC-ANN & Both \\
Nadiammai et al. '11 \cite{nadiammai2011comprehensive} & Misuse & KDD99 & ZR, DT, RF & Supervised \\
Lin et al. '12 \cite{lin2012intelligent} & Misuse & KDD99  & DT and SVM & Supervised \\
Kim et al. '14 \cite{kim2014novel} & Hybrid & NSL-KDD  & DT and SVM &  Supervised \\
De la Hoz et al. '14 \cite{de2014feature} & Hybrid & DARPA98, NSL-KDD &  NSGA-II+GH-SOM & Supervised\\
Eesa et al. '15 \cite{eesa2015novel} & Misuse &KDD99 & CFA-DT & Supervised\\
Lin et al. '15 \cite{lin2015cann} & Hybrid & KDD99 & Clustering & Supervised \\
Ravale et al. '15 \cite{ravale2015feature} & Hybrid & KDD99  & $k$-means + RBSVM & Supervised\\ 
Harshaw et al. '16 \cite{harshaw2016graphprints} & Anomaly & Self-made  & Graphprints & Unsupervised \\
Ikram \& Cherukuri '16 \cite{ikram2016improving} & Hybrid  & NSL-KDD, GURE-KDD   & PCA+SVM  & Supervised  \\
Mishra et al. '16 \cite{mishra2016efficient} & Hybrid & KDD99  & NB, NN, C4.5 DT &  Supervised\\
Zhu et al. '17 \cite{zhu2017intrusion} & Hybrid &  NSL-KDD, GURE-KDD, KDD99 & I-NSGA-III & Supervised \\
    \bottomrule

\end{tabular}
\end{table}

Barbar\'a et al.~\cite{barbara2001adam} introduce a new IDS, Audit Data Analysis and Mining (ADAM), for anomaly detection. 
ADAM is designed to be used on-line and is implemented using a data mining technique to build rules of normal behavior and a classifier to identify attacks in TCP dump recorded traffic. 
First, using data mining, ADAM builds a repository of attack-free items.
Then using an online, sliding-window algorithm, it identifies frequent items to compare with  the repository. 
Abnormal items are classified via a DT as a known attack, unknown attack, or false alarm. 
The system was tested using DARPA99 dataset and took third. 

Ambwani et al.~\cite{ambwani2003multi} propose a multi-class SVM classifier using a one-versus-one method\footnote{ To create an $n$-class classifier, the one-versus-one method trains binary classifiers to distinguish each of the  $n(n-1)/2$ pairs of classes. Usually a voting scheme collects the $n(n-1)/2$ votes to make a final classification. } to identify various misuse and attacks by type. 
The authors cite previous research to claim one-versus-one outperforms one-versus-all methods. 
An experiment is conducted using labeled 10\% training and testing sets from the KDD99 dataset. 
Accuracy (91\%) and performance are reported. 

Amor et al.~\cite{amor2004naive} compare NB and DT classifiers for intrusion detection, concluding DT is slightly more accurate but learning and classifying is seven times longer.
There were three experiments performed on the KDD99 traffic data, and both algorithms perform in the 91-94\% range. 

Liu et al.~\cite{liu2004genetic} describe an unsupervised, clustering-based approach, using a genetic algorithm to classify network traffic. This approach achieves good results overall when evalutated with the KDD99 datasets, and includes an adjustable sensitivity factor, i.e., a parameter to trade false positives for false negatives.

Chen et al.~\cite{chen2007hybrid} present an evolutionary NN\textemdash meaning the NN structure is learned as well as the parameters during training\textemdash trained by a particle swarm optimization (PSO) algorithm with flexible bipolar sigmoid activation functions. 
This work focuses on improving the intrusion detection performance by selecting an effective small subset of the input features. 
The algorithm's effectiveness is demonstrated on network data from the KDD99 dataset, and then its application to host-based datasets is discussed.

Kayacik et al.~\cite{kayacik2007hierarchical} demonstrate an unsupervised approach, using a hierarchy of SOMs, which results in similar detection rates as supervised approaches in other publications. 
When tested on the KDD99 dataset, the false positive rate of this unsupervised approach is about three times higher than supervised methods, and the authors discuss ways to reduce this further.

\"{O}zyer et al.~\cite{ozyer2007intrusion} describe an intelligent intrusion detection system (IIDS) that iteratively learns association and classification rules.  
Fuzzy association rule mining and filtering is used for each class, then a boosting algorithm is used for classification. 
Results appeared to be close to the KDD99 winning entry and proved the usefulness of this boosted fuzzy classifier. 
The authors believe that results can be farther improved with an increased training dataset.

Peddabachigari et al.~\cite{peddabachigari2007modeling} evaluate four classifiers for intrusion detection: DT, SVM, hybrid DT-SVM (first data is passed through a DT, and then the same data plus the result of the DT is passed through an SVM), and an ensemble, with the main goal to find the most accurate algorithm. 
Experiments are conducted on the KDD99 dataset, 
and it was determined that the ensemble approach performed the best. 
It classified non-attack, Probe, DoS, U2R, and R2U attacks with 99.70\%, 100\%, 99.92\%, 68\%, and 97.16\% of accuracy, respectively. 
To utilize the achieved results, a hybrid IDS is proposed, employing ensemble classifiers.

Tajbakhsh et al.~\cite{tajbakhsh2009intrusion} describe an IDS using fuzzy association rules and Association Based Classification (ABC) for misuse detection and an ABC extension for anomaly detection. 
The objective is a fast data mining technique that can be used to implement an IDS. 
The association rule induction algorithm grows exponentially according to the dataset's feature quantity. 
To accelerate this technique they introduce a concept called association hyper-edge, which reduces the above problem to a graph problem. 
The main challenge of the proposed classification is to define appropriate matching measures; two solutions are described and tested. 
Evaluations of their misuse detection approach using the KDD99 dataset realized a 91\% detection rate and a 3.34\% false positive rate; the anomaly detection approach had a detection rate of 80.6\% with 2.95\% false positive rate for all records with total execution time of about 500s. 
Detection of novel attacks was much less effective, as the misuse detection approach resulted in 11\% detection rate for new attacks, and the anomaly detection approach resulted in 20.5\% detection rate for these new attacks. 

Wang et al.~\cite{wang2010new} describe an IDS using fuzzy clustering (FC) to partition training data into subsets and train an NN on each subset;  
the results are aggregated with a fuzzy aggregation module for detection.   
An experiment was conducted using the KDD99 dataset, 
and results concluded that FC-ANN works similar to BPNN, NB, and DT in detecting high frequency attacks, and outperforms in detecting low frequency attacks. The average accuracy of this method is 96.71\%. Training time is higher than the other methods, but can be improved with parallelization. 

Nadiammai et al.~\cite{nadiammai2011comprehensive} determine the severity of attacks in a dataset with data mining methods. Several classifiers are used to determine the classification accuracy. 
The Zero R (ZR) classifier, DT classifier, and Random Forest (RF) classifier are compared using the KDD99 dataset. Results show that RF outperforms the other classification algorithms. 

Lin et al.~\cite{lin2012intelligent} develop an IDS based on SVMs and DTs, and use Simulated Annealing (SA)\textemdash a probabilistic optimization technique for avoiding local optima\textemdash for feature selection and parameter selection. This is evaluated with the KDD99 dataset, and the results show very high accuracy. SVM and SA can find the optimal subset of features for anomaly intrusion detection accuracy evaluation  while DT and SA can obtain decision rules for new attacks and can improve accuracy of classification.

Mishra et al.~\cite{mishra2016efficient} propose two IDS frameworks for a cloud network to detect a broad range of attacks and focus on speed and efficiency by employing parallelization techniques.  
Two types of parallelization were introduced, splitting traffic into multiple sensor nodes and using GPU acceleration programmed in CUDA. 
Experiments were conducted using the KDD99 dataset.  
The first setup used a single node
running an ensemble of NB, NN, and DT machine learning algorithms and resulted in 80-99.9\% of accuracy of each attack category. 
The authors note that NN requiress significant training time, so frequent retraining would likely be problematically time consuming. 
The second setup was done with the use of a multi-node environment for parallelization with 
each node running a single classifier: NB, NN, or DT.  
The second setup allows for a training time reduction for classifiers. 
The anomaly detection accuracy of DT and NN is about the same, ranging from 69.10-99.88\% for each attack category and NB produced the poorest result ranging from 45.20-98.30\%.

Harshaw et al.~\cite{harshaw2016graphprints} introduce a novel graph-mining and robust statistics technique to detect multi-scale (host and network level) anomalies. 
Self-harvested network flow data is used and a simple graph is constructed for each 31s time window with 1 second overlap between consecutive windows. 
The graphs' vertices represent an IP, and directed edges represent network flows, with two edge colors to encode port information. 
For network-level anomaly detection, small vertex-induced subgraphs, called graphlets, are counted\textemdash this records how many small subgraphs make up the whole network flow graph for each time window\textemdash and streaming anomaly detection is done on this sequence of vectors of graphlet counts. 
Anomalies in this sequence indicate the network has some local topological changes. 
For host-level detection, the host's role in an incident graphlet is counted (automorphism orbits), and anomalies in this sequence are flagged. 
The upshot of this method is that network-level anomalies can be pinpointed to the hosts that changed communication patterns. 
To perform the anomaly detetion on the sequence of vectors, a gaussian is fit using the Minimum Covariance Determinant method of Rousseeuw \& Driessen~\cite{rousseeuw1999fast}\textemdash this robust fitting algorithm automatically omits $h$\% of outliers and fits the gaussian distribution to the remaining inlier\textemdash with the goal of preventing attacks and other anomalies in the training (historical) data from affecting the models. 
Accurate detection of anomalous bit torrent traffic and port scans (although non malicous) were exhibited. 

Bridges et al.~\cite{bridges2017setting} focus on mathematical foundations for setting the threshold of probabilistic anomaly detectors. Their method can be considered in two ways\textemdash either one can set the threshold to bound the number of alerts (useful in high-throughput data situations to prevent flooding downstream systems) or if the alert rate bound is broken that indicates the probability model is a poor fit to the data. To exhibit the second alternative, Bridges et al. show that the robustly fit gaussians of Harshaw et al. are overfit to the inliers, i.e., $h=15$\% is too high for this application.


Many of the literature reviews found in this survey discuss detailed techniques for improved feature selection via reduction and construction. Additional feature selection algorithms for IDSs can be found in the survey by Chen et al.~\cite{chen2006survey}.

De la Hoz et al.~\cite{de2014feature} use the NSGA-II algorithm of Deb et al.~\cite{deb2002fast}, a multi-objective (maximize classification metrics while minimizing the number of features) feature selection algorithm with ``an unsupervised clustering procedure based on Growing Hierarchical Self-Organizing Maps'' (GH-SOMs) for both attack and anomaly detection.  
DARPA98 and NSL-KDD datasets used for evaluation. The selected feature sets are computed and reach 99.12\% accuracy while yielding normal and anomalous traffic detection rates as high as 99.8\% and 99.6\%, respectively.
Additionally, a reduction in the GH-SOM size improves the computational efficiency, making it a viable option for performing additional tasks (e.g., IP blocking) in real time.  

As non-attack data quantities far outnumber attack quantities, class imbalance is a perennial problem for machine learning intrusion detection algorithms. 
Zhu et al.~\cite{zhu2017improved} propose I-NSGA-III. 
Their algorithm is an improvement of NSGA-III, another multi-objective feature selection algorithm of Deb et al.~\cite{deb2014evolutionary}, to both alleviate redundant features and improve the accuracy in the face of imbalanced classes. 
Building on De la Hoz et al.~\cite{de2014feature}, 
they also use GH-SOM classifier for IDS and a similar method for feature selection. 
The algorithm can identify the attacks and distinguish between the different attack classes. 
The enhanced NSGA-III algorithm (I-NSGA-III) is compared to NSGA-II and NSGA-III algorithms among other algorithms. 
All are evaluated with NSL-KDD, KDD99, and GURE-KDD datasets. 
 
Results indicate that I-NSGA-III algorithm requires the least training and testing time with the best or near best detection rate.

Kim et al.~\cite{kim2014novel} introduce a new hybrid IDS for detection of anomalies and misuses. 
Their method is a combination of C4.5 DT and OCSVM methods. 
First, a DT is used for misuse detection and partitioning data into smaller subsets.  
Then, multiple OCSVMs are applied to each subset for anomaly detection. 
Using the NSL-KDD dataset for evaluation, results indicate that the proposed hybrid IDS has higher detection rate by about 10\%, while false positive rate is below 10\%, 
and shows faster training compared to other conventional methods. Training and testing time was 56.58s and 11.20s, respectively.

Ravale et al.~\cite{ravale2015feature} create a hybrid learning approach by first applying $k$-means clustering, to reduce the large heterogeneous dataset into multiple, smaller homogeneous subsets, and to select features.  Following this, a Radial SVM classifier is used with the selected features. 
Evaluation with the KDD99 dataset resulted in increased performance with greater rates of accuracy and detection, while achieving a fewer false alarms than either SVM or $k$-means independently.

Eesa et al.~\cite{eesa2015novel} describe an IDS approach that minimizes the quantity of features utilized while maximizing the detection rate. 
For feature-selection, a new cuttlefish optimization algorithm (CFA) \cite{eesa2013cuttlefish} is used as a search method to determine the ideal feature subset, and then a DT uses the CFA-selected subset of features to perform classification. 
An experiment was conducted using the KDD99 dataset; this was processed eight times through the DT classifier with different subsets of of the CFA-selected features, ranging from 5 to 41 features. The results reveal that the feature subsets containing 20 features or less, ascertained with CFA, provide greater rates of both detection and accuracy while presenting fewer false alarms than results utilizing every feature. 

Lin et al.~\cite{lin2015cann} describe a feature representation method that combines Cluster Centers and Nearest Neighbors (CANN). 
The CANN method converts original features into a single  ``distance-based feature'' that is fed to a $k$-NN classifier. 
This gives an initial computation cost for the dimension reduction, but yields better performance in detection. 
Results on KDD99 data show that CANN yields better accuracy, and true positive and false positive rate than either $k$-NN or SVM classifiers with unmodified six features. 
CANN is, however, ineffective at both U2R and R2L attack detection. 

Ikram \& Cherukuri~\cite{ikram2016improving} describe a hybrid IDS with improved accuracy and lower resource consumption, using principal component analysis (PCA) for initial dimension reduction followed by an optimized SVM. 
The novelty of this proposed approach is optimization of an SVM punishment factor and RBF kernel's gamma parameter using automatic parameter selection. 
To test this proposed approach, NSL-KDD and GURE-KDD network traffic datasets were used. 
Detection accuracy for NSL-KDD data is reported before and after applying PCA of each class\textemdash Probe, DoS, U2R, R2L and Normal classes\textemdash  from 85.65\% to 90.84\%, from 46.41\% to 80.43\%, from 33.33\% to 78.35\%, from 87.19\% to 78.35\% and from 97.38\% to 63.62\%,  respectively. 

We conclude this subsection with an introspective work on the flurry of algorithmic research spawned by the KDD and DARPA datasets. 
Motivated by many machine learning algorithms achieving greater than 90\% detection rates with less than 1\% false positive rates  using the KDD99 dataset \cite{pfahringer2000winning, levin2000kdd, eskin2002geometric, kayacik2003capability}, Kayacik et al.~\cite{kayacik2005selecting} conduct feature-relevance analysis to gain insight. 
The most relevant features with respect to dataset's labels are determined by employing information gain in conjunction with DTs; therefore, the highest information gain determines the most discriminative features, which makes classification easier. 
Due to the need for distinct features for information gain calculation, continuous features are converted to discrete features by equally partitioning with the equal frequency intervals method \cite{wong1987synthesizing}. 
Three of the provided classes, ``normal", ``smurf", and ``neptune", are highly related to certain features and are comprised of 98\% of the training data thus making it easily classifiable for machine learning algorithms. 
While the quantity of data exchanged is the most discriminating feature for 14 of the 23 classes, some features have no effect on intrusion detection, indicating that they are not needed.

\section{Conclusion}
\label{sec:conclusion}
This survey provides an overview of IDS types including various locations and types of IDSs. 
Related surveys are identified, and focus is given to HIDSs. 
In order to organize works and itemize the available data sources, 
the HIDS literature is presented per input data source.  
In particular, system logs, audit data, Windows Registry, file systems, and program analysis detection works are sub-categories investigated. 
Specific sections are allocated for system call IDS and another for algorithmic research tested on network-level data but applicable to host data. 
A large number of works fall into these two categories because of the publicly available labeled datasets with these types of data.
We conclude with a subsection outlining limitations and budding directions for HIDS research. 
Additionally, this survey compiles a supplementary list of many publicly available datasets, with descriptions of their characteristics and shortcomings.

\subsection{Suggested Future Directions} 
Although there is a wealth of IDS literature, successfully transitioned-to-practice HIDS techniques are rare, with OCSEC and Tripwire as outstanding counter examples. 
This is due to a number of factors that point to directions for future progress in HIDS research.

First, IDS research is constrained by limited available datasets and ``in vitro'' development (where test environments fail to capture the complexities of real networks). 
For many data sources, there are either no datasets publicly available, or those that are available are outdated, low-quality, lacking in attack diversity, or contain other serious flaws. 
This leads to researchers often simulating or otherwise building datasets that often lacks fidelity, complexity, or realistic benign activity.  
Alternatively, when real data is recorded and used, it is generally not sharable due to privacy or security concerns, and may still contain many of the limitations above (e.g. lack of attack diversity.)
As evidenced by the explosion of IDS research spawned by the few well-adopted datasets (DARPA, KDD, UNM, ADFA, most notably), 
these facilitate quantifiable comparison of techniques across publications and provide accessible data to the hands of eager researchers, in spite of their many flaws. 
Up-to-date efforts to curate and publicize realistic, attack-labeled, and ideally multi-source host and network data sets will likely be met with a similarly large response from the IDS research community. 
Moreover, for supervised learning techniques,  addressing the question of training for actual operational use is a necessity, e.g., providing a validated method for generating training data that combines a real host's/network's data with labeled attacks.

Second, HIDS research is preoccupied with (admittedly important) detection metrics at the expense of understandability of alerts. 
Indeed, for adoption of an HIDS, gaining the user's trust in terms adequate testing to establish an acceptable true-to-false positive balance is a necessity. 
On the other hand, the myriad of publications that flex their statistical prowess and claim success upon incriminating detection rates often fail to provide actionable results to the operator. 
This ``semantic gap'' problem is perhaps first established by the famous Sommer \& Paxson work \cite{sommer2010outside}. 
Security information and event management (SIEM) tools, which correlate alerts and logs from diverse systems in real-time to enhance operators understanding, are emerging in the commercially available tools, and research providing open-source options also are developing, e.g., see Stucco\footnote{ See \url{https://github.com/stucco}.}.  
Research is needed to leverage the many diverse but related data sources available to an HIDS, (not only to increase detection accuracy, but) to provide a contextual, situational awareness along with an accurate alert is needed to operationalize much of the work surveyed here.

We note that the new Unified Host and Network Data set of Turcotte et al. \cite{turcotte17} contributes to these first two directions by providing real network and host data for researchers. 
Further, efforts such as Ilgun~\cite{ilgun1993USTAT} provide an automated component to present the alerts to the user in a smart way.

Finally, while many researchers provide adequate investigations of the computational burden of their IDS,  this is a known inhibitor of HIDS deployment. 
Research to dynamically change the IDS for dual optimization of increased security and decreased overhead is needed.  
Examples may include dynamic algorithms to adjust detector alert thresholds, change computational requirements, adjust data sources collected, or change the position or security posture of the host, based on current conditions to provide a more effective tradeoff between resources and security. 
Some works have begun these investigations, in particular, for cloud applications \cite{lee2011multi} and for threshold tuning~\cite{bridges2017setting}.

Overall, we hope our treatment of the HIDS literature provides an organized panorama for researchers to gain insights, identify opportunities, and more quickly progress in advancing HIDSs.

\section*{Acknowledgement}
The authors would like to thank the many reviewers who have helped polish this document, in particular, Jared M. Smith.
The authors also thank Kerry Long for fruitful conversations contributing to this work's scope and direction. 
The research is based upon work supported by the Office of the Director of National Intelligence (ODNI), Intelligence Advanced Research Projects Activity (IARPA), via the Department of Energy (DOE) under contract  D2017-170222007. 
The views and conclusions contained herein are those of the authors and should not be interpreted as necessarily representing the official policies or endorsements, either expressed or implied, of the ODNI, IARPA, or the U.S. Government. 
The U.S. Government is authorized to reproduce and distribute reprints for Governmental purposes notwithstanding any copyright annotation thereon. 


\small

\bibliographystyle{abbrv}
\bibliography{refs}

\section*{Supplementary Section: Datasets}
\addcontentsline{toc}{section}{Supplementary Material}
\label{sec:datasets}


Intrusion detection evaluation datasets are important resources for validation, comparison, and experimentation. 
Popularity of a labeled dataset among researchers allows comparison of detection metrics or performance across publications, and in many cases has stimulated a flurry of IDS research on a particular data source. 
Common pitfalls of such datasets are artificial artifacts correlated with targets, unrealistic attacks, and redundant or missing data among others. 
Here we have compiled the datasets commonly used in the HIDS research literature with a brief description of their contents, and noteworthy advantages or drawbacks.  
Table~\ref{tab:datasets} gives itemized information at a glance, and the website for each data source is at the conclusion of its description in the text.

\begin {table}[!ht]
\small
\caption {Datasets and Public Datasets and Dataset Collections} 
\label{tab:datasets} 
\begin{tabular}{p{.2\textwidth} p{.14\textwidth} p{.18\textwidth} p{.4\textwidth} }
    \toprule
      Name/Abbreviation & Data Source  & Attack Class & Website\\
\midrule
\small
IMPACT & NA & Various  & \url{http://www.impactcybertrust.org} \\
\mbox{Digital Corpora} \mbox{Database} & NA &  Various & \url{http://digitalcorpora.org} \\
DARPA98, 99, 00 & Network traffic, System calls & 
\mbox{DOS, U2R, R2L,} PROBE & \url{http://www.ll.mit.edu/mission/communications/cyber/CSTcorpora/ideval/data} \\
KDD99 & Network traffic & \mbox{DOS, U2R, R2L,} PROBE & \url{https://kdd.ics.uci.edu/databases/kddcup99/kddcup99.html} \\
NSL-KDD & Network traffic  & \mbox{DOS, U2R, R2L,} PROBE & \url{http://www.unb.ca/cic/research/datasets/nsl.html} \\
GURE-KDD & PCAPs & \mbox{DOS, U2R, R2L,} PROBE & \url{http://aldapa.eus/res/gureKddcup}\\
UNM & System calls & \mbox{Buffer overflows,} \mbox{symbolic link,} trojans &
\url{http://www.cs.unm.edu/~immsec/systemcalls.htm}\\
ADFA-LD, ADFA-WD, ADFA-WD:SAA &  System calls & Exfiltration, DDoS, other  & \url{https://www.unsw.adfa.edu.au/australian-centre-for-cyber-security/cybersecurity/ADFA-IDS-Datasets}\\
Active DNS Project & DNS PCAPs &  \mbox{Malware, spam,} phishing, other &  \url{https://www.activednsproject.org}\\
SecRepo & malware, NIDS,  \mbox{host logs,} PCAPs& Various & \url{http://www.secrepo.com} \\
\mbox{Malware Traffic} \mbox{Analysis} & Malware, PCAPs & Malware & \url{http://www.malware-traffic-analysis.net}\\
NETRESEC & PCAP DBs list & Various & \url{http://www.netresec.com/?page=PCAPFiles} \\
CTU 13 & Network flow, PCAPs  & Botnet & \url{https://goo.gl/i9WQq3} \\
\mbox{Malware Capture} \mbox{Facility} Project & PCAPs & Botnet, Various & \url{https://www.stratosphereips.org/datasets-malware} \\
The Honeynet Project & Malware, PCAPs, logs & Various & \url{http://honeynet.org/challenges} \\
VAST Challenge 2013 & Network flows, logs  & \mbox{DOS, FTP exfil.,} other & \url{http://vacommunity.org/VAST+Challenge+2013} \\
VAST Challenge 2012 & Network logs  & \mbox{Botnet, scanning,} exfil. & \url{http://vacommunity.org/VAST+Challenge+2012} \\
UNSW-NB15 & PCAPs & Fuzzers, backdoors, \mbox{DoS, exploits,} \mbox{recon}, other & \url{https://www.unsw.adfa.edu.au/australian-centre-for-cyber-security/cybersecurity/ADFA-NB15-Datasets/} \\
CAIDA & PCAP headers, \mbox{other internet} data & Unlabeled  & \url{http://www.caida.org/data} \\
\mbox{Unified Host and} \mbox{Network} Dataset & \mbox{Network \&} host audit data &  Unlabeled & \url{https://csr.lanl.gov/data/2017.html} \\

    \bottomrule
\end{tabular}
\end{table}

\begin{description}[leftmargin = .4cm] 
\item[Information Marketplace for Policy and Analysis of Cyber-Risk \& Trust (IMPACT)] 
The Department of Homeland Security maintains the IMPACT database\footnote{ See \url{https://www.impactcybertrust.org}.}. Formerly known as PREDICT, the ``Protected Repository for the Defense of Infrastructure Against Cyber Threats'', IMPACT contains recent network operations data contributions from developers around the world aiming to improve cyber-risk research and development. The cyber-related dataset repository is publicly available. 

\item[Digital Corpora] Computer forensics education research data including disk images, memory dumps, network packet captures, etc., is available freely in this database. 
Additionally, Digital Corpora provides a research corpus of real data acquired from around the world; however, usage is limited.

\item[DARPA Intrusion Detection 1998, 1999, \& 2000] The ``Cyber Systems and Technology Group'' of MIT Lincoln Laboratory, working with DARPA (Defense Advanced Research Projects Agency) and AFRL (Air Force Research Laboratory), created the first public, standard corpora intended for evaluation of computer network intrusion detection systems~\cite{Darpa}. 

The 1998 dataset is a widely-used collection of known attacks, and consists of system call based audit data and network data, including full packet capture.
The data is comprised of \texttt{praudit}\footnote{ See \url{https://docs.oracle.com/cd/E19253-01/816-4557/auditref-76/index.html} for description of the \texttt{praudit} command.} and \texttt{list} files, as well as packet captures from \texttt{tcpdump}; 
The attacks conducted to generate this dataset were not automated, 
and they are considered high footprint attacks by subsequent researchers.
~\cite{haider2015integer}

Numerous issues have been documented with this dataset \cite{mchugh2000testing, mahoney2003analysis, brugger2007assessment}. 
Brugger and Chow~\cite{brugger2007assessment} noted several issues with the dataset including inability to accommodate the latest attack trends, 
and the majority of malicious connections 
consisting of denial of service attacks and probing activity. 


The 1999 dataset consists of a series of network packet dumps and BSM system call records. The data has been widely used in the intrusion detection and networking community, even though it is known to have a number of artifacts of its creation, including the lack of damaged or unusual background packets and uniform host distribution.~\cite{mahoney2003analysis}

\item[KDD Cup 1999 (KDD99)] This dataset was created by processing the \texttt{tcpdump}\footnote{ See \url{https://danielmiessler.com/study/tcpdump/} for a tutorial on the \texttt{tcpdump} command/tool.}  portions of the DARPA98 dataset. It provides labeled data for intrusion detection and contains four attack types; DoS (denial of service), U2R, R2L, and PROBE~\cite{pachghare2012pattern}. 
However, evaluating machine learning algorithms, such as DTs \cite{pfahringer2000winning, levin2000kdd}, NNs \cite{kayacik2003capability}, and SVMs \cite{eskin2002geometric}, with KDD99 substantiates that it is not possible to accurately detect U2R and R2L attacks. 
Sabhnani \& Serpen~\cite{sabhnani2004machine} investigated the KDD dataset deficiencies and concluded that for the U2R and the R2L attack categories no trainable pattern classification or machine learning algorithm can achieve an acceptable level of misuse detection performance on the KDD testing data subset if classifier models are built using the KDD training data subset. This is due to the omission of attacks and their records from the training data subset. 

\item[NSL-KDD] The NSL-KDD dataset was created to improve upon the shortcomings of the KDD99 dataset.
KDD99’s record redundancy hinders an algorithm’s ability to learn by causing a bias against infrequent records and, in turn, overlooking harmful attacks. This issue was resolve with the removal of duplicate records in both the training and testing sets. 
Consequently, the reduction makes it feasible to run the experiments on the full set without requiring random subset selection. .~\cite{tavallaee2009detailed}

\item[GureKddcup and GureKddcup6percent] GureKddcup consists of the KDD99 connections with added network packet payload that allows for direct extraction by learning algorithms. The GureKDDcup dataset is generated by following the same steps as the KDD99 dataset and consists of numerous redundant entries. Bro-IDS is used for processing the \texttt{tcpdump} files to acquire connections along with their attributes. All connections are labeled with MIT's ``connections-class'' files~\cite{sahu2014detail}. The original dataset size is 9.3GB, and the 6\% dataset size is 4.2GB.~\cite{gurekdd}

\item[University of New Mexico dataset (UNM)] In 2004, the UNM dataset was released consisting of four datasets of systems calls executed by active processes; ``Synthetic Sendmail UNM, Synthetic Sendmail CERT, live lpr UNM, and live lpr MIT''~\cite{elgraini2012host}. 
Several programs are included ``(e.g., programs that run as daemons and those that do not), programs that vary widely in their size and complexity, and different kinds of intrusions (buffer overflows, symbolic link attacks, and Trojan programs).’’~\cite{UNM}. The dataset consists of both ``synthetic'' and ``live'' traces, and a trace consists of a list of a unique process’ system calls. The UNM dataset is as antiquated as the KDD data and focuses on individual processes rather than the entire OS~\cite{creech2013generation}.

\item[ADFA IDS datasets (ADFA-LD, ADFA-WD and ADFA-WD:SAA)] Since performance on the Darpa98 and KDD99 datasets does not represent true performance against contemporary attacks, ADFA was developed as a modern benchmark for HID. 
The ADFA IDS labeled dataset is the successor of the KDD collection using the latest publicly available exploits and methods. 
There are three groups of data with raw system call traces: training, testing normal, and testing attack.
The dataset is designed for use with an anomaly based IDS so there are no attack traces used during training. 

All training and validation data traces were gathered under normal host operations, during activities varying from browsing the web to \LaTeX document generation. The ADFA dataset contains more similarities between attack data and normal data than either the Darpa98 or the KDD99 datasets. This allows for a more accurate portrayal of cyber attacks and better assessment of IDS performance.~\cite{creech2013generation}

Two Windows OS specific datasets were generated to protect from zero-day attacks, stealth attacks, data exfiltration, and DDoS attacks. 
ADFA-WD is comprised of known ``Windows based vulnerability oriented zero-day attacks'' and ADFA-WD:SAA is an expansion used for resistance validation of prospective HIDS.~\cite{haider2016windows}

  
\item[Active DNS Project] Over a terabyte of ``unprocessed DNS packet captures'' (PCAPs) along with a plethora of daily de-duplicated DNS records.~\cite{kountouras2016enabling}. 
  
\item[Security Repo (SecRepo)] The SecRepo is a compilation of Security data including malware, NIDS,  Modbus, and system logs. Additionally, it consists of several of the following datasets.
  
\item[Malware Traffic Analysis] Samples of malware binaries and PCAPs are provided along with an active campaign listing. \footnote{ See \url{http://www.malware-traffic-analysis.net/2018/index.html}.} 

\item[NETRESEC Data] This data is a list of public packet capture repositories, which are freely available on the Internet. Most of the sites listed below share Full Packet Capture (FPC) files, but some do unfortunately only have truncated frames. This includes SCADA/ICS Network Captures.
  
\item[CTU 13] The data contains 13 datasets \footnote{ See \url{https://mcfp.weebly.com/the-ctu-13-dataset-a-labeled-dataset-with-botnet-normal-and-background-traffic.html}}, each containing a malware binary, a network flow \texttt{.csv} file from the ARGUS flow sensor \footnote{ See \url{http://qosient.com/argus}.}, and PCAP file(s) with botnet traffic. 
Included in every dataset is a \texttt{readme} file providing information for which IPs are infected or attacked and how.~\cite{garcia2014empirical}
    
\item[Malware Capture Facility Project] This dataset is an extension of the CTU 13 dataset, and consists of the similar information from around 350 attacks pertaining to malicious PCAPs.


\item[The Honeynet Project] Consists of a variety of data from all of the challenges, including PCAP, malware, and logs.

\item[VAST Challenge 2013 Mini-Challenge 3] This is a cybersecurity challenge that includes data related to network flow, network status, and intrusion prevention systems. However, there are sizable data gaps.

\item[VAST Challenge 2012] This challenge consists of two smaller tasks. The first involving situational awareness (e.g., metadata and periodic status reporting) and the second involving forensics (e.g., Firewall and IDS logs).



\item[UNSW-NB15] A comprehensive dataset for NIDS containing nine attacks types: Fuzzers, Analysis, Backdoors, DoS, Exploits, Generic, Reconnaissance, Shellcode, and Worms. The ARGUS flow sensor and Bro-IDS\footnote{ See \url{https://www.bro.org/sphinx/broids/index.html}.} tools are used along with the development of twelve algorithms for the generation of 49 features with the class label.

\item[Center for Applied Internet Data Analysis (CAIDA)] 
The CAIDA Anonymized Internet Traces 2016 annual dataset consists of anonymized traffic traces with a single trace generated quarterly.  The internet traffic contains ``application breakdown, security events, topological distribution, and flow volume and duration.'' Software capable of reading packet captures (PCAPs) in  \texttt{tcpdump} format can read the traces. All traces are made anonymous with the same key using ``CryptoPan prefix-preserving anonymization’’ and there is a complete packet payload removal. 
There is a negligible quantity of packet lost for some data.~\cite{caida} 

\item[Unified Host and Network Dataset] The ``Unified Host and Network Dataset'' consists of both network and host event data gathered from Los Alamos National Laboratory's (LANL) over approximated ninety days. 
The host event logs come from Microsoft Windows OS machines and the network event data comes from ``router network flow records.'' Although there is overlap in the Windows OS machines use for both the network and host datasets, the network dataset also utilizes additional machines running other OSs.~\cite{turcotte17}

\end{description}

\begin{wraptable}[10]{R}{4.5cm}
\vspace{-1.1cm}
\centering
\caption{DARPA Attack Classes}
\label{attackClasses}
\begin{tabular}{|c|l|}
\hline
\rowcolor{LightCyan} 
\multicolumn{1}{|l|}{\cellcolor{LightCyan}\textbf{Attack Class}} & \textbf{Class Name} \\ \hline
0                                                                   & Normal              \\ \hline
1                                                                   & Probe               \\ \hline
2                                                                   & DoS                 \\ \hline
\rowcolor{LightGrey}
\multicolumn{2}{|c|}{\cellcolor{LightGrey}\textbf{Compromises}}                        \\ \hline
3                                                                   & U2R                 \\ \hline
4                                                                   & R2L                 \\ \hline
\end{tabular}
\end{wraptable} 

\subsection{Common Attack Types in Publicly Available Datasets}

The records in the DARPA- and KDD-related datasets include attack types and can be classified into one of five classes: Probe, DoS, U2R, R2L, and Normal. 

Many papers included in this survey refer to the traditional attack classes by the numbering convention provided in Table \ref{attackClasses}. The table and definitions provided below can be used as a quick reference.

The last two classes are considered compromises and occur when an attacker gains privileged access to host access after hacking into the system through insecure points. Compromises are separated into two classes depending based on the source of the attack.

\begin{enumerate}[leftmargin = .4cm]
\item Probing (surveillance, scanning): Attacker tries to gain information about the target host, e.g., port scanning. These attacks collect lists of potential vulnerabilities through network scans that can be utilized later in an attack against the machine or service.

\item Denial of Service (DoS): Attacker tries to prevent legitimate users from using a service, e.g., using SYN flood. These attacks an occur on both the operation system; targeting bugs, or in the network; exploiting protocols and infrastructure limitations. 

\item User to Root (U2R): The attack is derived from within the system. An attacker who has local access to the victim machine tries to gain root access by exploiting a vulnerability, e.g., local buffer overflow attacks.

\item Remote to Local (R2L): The attack is derived from outside the system, over the network. The attacker does not have access to any legitimate account on the victim machine, therefore tries to gain access. This is commonly achieved through the Internet using password guessing attacks or exploits allowing remote code execution. 

\end{enumerate}

\end{document}